\documentclass[seceq]{ptptex}

\usepackage{graphicx}
\usepackage{wrapft}
\usepackage{amsmath,amssymb}

\title{
Bosenova collapse of axion cloud
around a rotating black hole%
}

\author{
Hirotaka \textsc{Yoshino}$^{1}$
and Hideo \textsc{Kodama}$^{1,2}$
}

\inst{
$^1$Theory Center, 
Institute of Particles and Nuclear Studies, 
KEK, Tsukuba, Ibaraki, 305-0801, Japan\\
$^2$Department of Particle and Nuclear Physics,  
Graduate University for Advanced Studies, Tsukuba 305-0801, Japan
}

\abst{
Motivated by possible existence of
stringy axions with ultralight mass,
we study the behavior of an axion field
around a rapidly rotating black hole (BH)
obeying the sine-Gordon equation
by numerical simulations.
Due to superradiant instability,
the axion field extracts 
the rotational energy of the BH and  
the nonlinear self-interaction
becomes important as the field grows larger.
We present clear numerical evidences that the nonlinear effect
leads to a collapse of the axion cloud and a 
subsequent explosive phenomena,
which is analogous to the ``bosenova'' observed
in experiments of Bose-Einstein condensate. 
The criterion for the onset of the bosenova collapse
is given. We also discuss the reason why the bosenova happens
by constructing an effective theory of 
a wavepacket model 
under the nonrelativistic approximation. 
}

\begin{document}

\maketitle

%
%
\section{Introduction}

Recently, it was pointed out that string theory 
may be probed through cosmology or astrophysics by
observing phenomena caused by 
``axions'' \cite{Arvanitaki:2009,Arvanitaki:2010}
(see Ref.~\citen{Kodama:2011} for a review).
The axion usually refers to the QCD axion that was introduced 
to solve the strong CP problem by Peccei and Quinn 
\cite{Peccei:1977-1,Peccei:1977-2}, and
the QCD axion is expected as one of the candidates of the
dark matter. In addition to the QCD axion, in the context of string theory,  
the stringy axions or axionlike particles have been
proposed and discussed \cite{Arvanitaki:2009,Arvanitaki:2010}. 
In string theory, many moduli 
arise when extra dimensions are compactified, and some of them
are expected to behave as axionlike scalar fields 
with ultralight mass. The typical
expected number of the axionlike particles is 10--100, and it leads to
a generic landscape of stringy axions, the so-called 
``{\it axiverse}''.

The ultralight axions cause the possibly observable phenomena
in cosmology or astrophysics. 
Suppose that the decay constant $f_a$ of the
axion is order of the GUT scale, $f_a\approx 10^{16}$GeV. 
In the cosmological context, 
axions with mass from $10^{-33}\mathrm{eV}$ to 
$4\times 10^{-28}\mathrm{eV}$
affect the polarization of cosmic microwave background
if the Chern-Simons interaction is present, and those with mass
from $4\times 10^{-28}\mathrm{eV}$ to $3\times 10^{-18}\mathrm{eV}$
may affect the matter power spectrum. 
On the other hand, in the astrophysical context, 
axions are expected to cause interesting
phenomena around astrophysical black holes (BHs) if they have mass between 
$2\times 10^{-20}\mathrm{eV}$
and $3\times 10^{-10}\mathrm{eV}$. 
We focus on axions around an astrophysical BH
in this paper.

Suppose an axion field exists around a rotating BH.
Although some of the field would be absorbed by the BH, 
it is expected that the axion field forms a quasibound state
which may be called the ``axion cloud''.
Furthermore, the axion cloud is expected to grow 
by the superradiant instability. 
The superradiant instability is 
caused by the fact that the Killing vector field $(\partial_t)^a$
of the Kerr spacetime 
becomes spacelike in the ergoregion, and therefore,
the energy of the field can be negative.
A mode with negative energy around the horizon
is called a superradiant mode. 
If a quasibound state whose mode function 
satisfies the superradiant condition 
is occupied by the axion field, 
negative energy falls into the BH and the energy outside 
the horizon (and therefore, the amplitude of the field) increases in time.

The growth rate of the quasibound state of the axion cloud
by the superradiant instability 
is characterized by the imaginary part $\gamma$ of the angular
frequency $\omega$, where $\omega = \omega_0 + i\gamma$. 
The value $\gamma/\mu$ depends on the ratio 
of one half of the gravitational radius to the Compton wavelength of
axion $\alpha_g:=(GM/c^2)/(\hbar/\mu c)$, or, $\alpha_g=M\mu$
in the Planck units $c=G=\hbar=1$ 
(hereafter, the Planck units are used unless otherwise specified). 
The superradiant instability
is effective for $\alpha_g\sim 1$, and 
its typical time scale is $\tau\sim 10^{7}M$ for $\alpha_g\sim 1$. 
For a solar-mass BH $M=M_{\odot}$
(resp. a supermassive BH $M=10^{9}M_{\odot}$), 
the superradiant instability effectively occurs
if an axionlike field with mass $\mu \sim 10^{-10}\mathrm{eV}$
(resp. $\mu \sim 10^{-19}\mathrm{eV}$) exists, and in that case, the typical
time scale for the instability is $50 \ \mathrm{s}$ 
(resp. $1600 \ \mathrm{year}$).
Therefore, the time scale for instability is much shorter
than the age of the universe and 
the superradiant instability should become 
really relevant to astrophysical phenomena.

The expected phenomena caused by the superradiant instability are
discussed and summarized in 
Refs.~\citen{Arvanitaki:2009,Arvanitaki:2010}\footnote{This 
problem was also discussed by a different method in
Ref.~\citen{Mocanu:2012}.}.
As the instability proceeds, 
the axion cloud extracts rotational energy from a BH
and gradually becomes heavy (i.e., the number of axions increases). 
In Ref.~\citen{Arvanitaki:2010}, the gravitational wave emission
was discussed from the viewpoint of the quantum theory. 
Since the structure of the axion cloud is analogous to the
electron cloud of a hydrogen atom, the
graviton emission by the level transition of axions can be discussed
in analogy with the photon emission by electron's level transition.
Another source of graviton emission is the pair annihilation
of two axions.  
On the other hand, the nonlinear self-interaction of axions
is also expected to cause important phenomena. 
In the case of the QCD axions, due to nonperturbative effects
associated with instantons, 
the potential $U(\Phi)$ becomes periodic as typically described by 
the trigonometric function
$U(\Phi) = f_a^2\mu^2[1-\cos(\Phi/f_a)]$.
The similar form of the potential can be expected 
for string axions because their masses
are generated also by the instanton effects.
Therefore, although the Klein-Gordon equation 
(i.e., $U(\Phi) = (1/2)\mu^2\Phi^2$)
gives a good approximation for small $\Phi/f_a$,
as the field grows large, the nonlinear effects become important.

One of the nonlinear effects is the mode mixing, which
is expected to change the field configuration and
affect the growth rate of the superradiant instability.
Another interesting possibility is the ``bosenova''.
The bosenova was observed in the experiments of 
the Bose-Einstein condensates (BEC) of Rb85 \cite{Cornish:2000,Donley:2001}. 
The interaction between atoms can be controlled
in this system, and the interaction was switched
from repulsive one to interactive one in that experiment. 
As a result, the BEC collapsed, but 
after that, a burst of atoms was observed. This phenomenon 
was studied also theoretically \cite{Saito:2001-1,Saito:2001-2,Saito:2002} 
and it was clarified 
that the implosion is caused by the nonlinear attractive interaction
and the burst is induced mainly 
by atomic loss through three-body recombinations.
Since the atomic loss weakens the attractive interaction, the atoms
begin to explode due to zero-point kinetic pressure.

In the case of the BH-axion system, we have to
take account of the following two possibilities. The first possibility is that
an explosive phenomena that is analogous to the bosenova happens as a
result of the nonlinear effect. The second possibility 
is that the nonlinear effect saturates
the growth by superradiant instability, as found in various 
instabilities of nonlinear systems, leading the system to 
a quasistationary state without explosive phenomena. 
We have to clarify which is the case, and 
if phenomena like the bosenova 
also occurs in the BH-axion system,  
the details and the observational consequence have to be studied. 
In order to clarify
the strongly nonlinear phenomena, 
fully nonlinear simulations have to be performed, and
this is the purpose of this paper.

We develop a three-dimensional (3D) code to simulate an 
axion field with a nonlinear potential 
in a Kerr spacetime. Here, the axion field is treated
as a test field, and the background geometry is fixed
to be the Kerr spacetime. In most cases, this approximation
holds well. The setup of the problem is explained in more detail
in Sec.~\ref{Sec:setup}. In short, our simulations indicate
no evidence for saturation, and the bosenova is likely to 
happen in the final stage of the superradiant instability.

This paper is organized as follows. 
In the next section, we review the existing studies on 
the behavior of a massive scalar field and its 
superradiant instability focusing attention to the
aspects closely related to our study. 
Section~\ref{Sec:Techniques} explains the technical part,
i.e., the formulation, our code, and code tests.
In Sec.~\ref{Sec:results}, we present the numerical results
of our simulations. After presenting the results of typical
two simulations, we discuss whether the bosenova actually
happen by performing supplementary simulations.
In Sec.~\ref{Sec:effective_theory}, we discuss the reason
why the bosenova happens in the BH-axion system
by constructing an effective theory of an axion cloud model
in the nonrelativistic approximation. 
Section~\ref{Sec:summary} is devoted to summary and discussion.
After summarizing our results, the similarity and difference
between the bosenova phenomena in the BEC system and in the
BH-axion system is discussed. 
We also roughly estimate whether gravitational radiation emitted 
in the bosenova can be detected
by planned gravitational wave detectors.
In Appendix A, the behaviour of the axion field 
generated by the nonlinear effect is studied 
using the Green's function approach, taking attention to the
consistency with the results of our simulations.

%
%
\section{Superradiant instability}
\label{Sec:Superradiant_instability}

This section is devoted to the review 
on massive scalar fields in a Kerr spacetime.

\subsection{Axion field in a Kerr spacetime}
\label{Sec:Field_Kerr}

The action for the axion field $\Phi$ 
in a spacetime of a metric $g_{ab}$ is
\begin{equation}
S=\int d^4x\sqrt{-g} \left[-\frac12
g^{ab}\nabla_a\Phi\nabla_b\Phi
-U(\Phi)
\right],
\label{Eq:action}
\end{equation}
where $U(\Phi)$ is the potential, i.e., $U(\Phi) = (1/2)\mu^2 \Phi^2$
for the Klein-Gordon field and $U(\Phi) = f_a^2\mu^2[1-\cos(\Phi/f_a)]$
for the axion field with nonlinear self-interaction (i.e.,
the sine-Gordon field). Here,
$f_a$ is the decay constant whose value depends on the model.
For convenience, we  normalize the amplitude of $\Phi$ with $f_a$ as
\begin{equation}
\varphi:= \Phi/f_a.
\label{Eq:def_varphi}
\end{equation}
Then, the field equation is  
\begin{equation}
\Box \varphi - \hat{U}^\prime (\varphi) = 0.
\label{Eq:field_equation}
\end{equation}
with $\hat{U}(\varphi) = U(\Phi)/f_a^2$. 
Here, 
$\hat{U}^\prime = \mu^2\varphi$ for the
Klein-Gordon field and $\hat{U}^\prime = \mu^2\sin\varphi$
for the axion field. Therefore, if 
the value of $\left|\varphi\right|$ is sufficiently small,
the axion field can be well approximated by the Klein-Gordon field.
However, the nonlinear effect becomes important 
as $|\varphi|$ comes close to unity.

The metric of the Kerr spacetime in the Boyer-Lindquist 
coordinates is given by
\begin{multline}
ds^2 = -\left(\frac{\Delta - a^2\sin^2\theta}{\Sigma}\right)dt^2
-\frac{2a\sin^2\theta(r^2+a^2-\Delta)}{\Sigma}dtd\phi\\
+\left[\frac{(r^2+a^2)^2-\Delta a^2\sin^2\theta}{\Sigma}\right]
\sin^2\theta d\phi^2
+\frac{\Sigma}{\Delta}dr^2
+\Sigma d\theta^2,
\end{multline}
with
\begin{equation}
\Sigma = r^2+a^2\cos^2\theta, \qquad \Delta = r^2+a^2-2Mr.
\end{equation}
Here, $M$ is the Arnowitt-Deser-Misner (ADM) mass, and 
$a$ is the ADM angular momentum per unit mass, $a=J/M$.
In order to specify the rotation, 
the nondimensional parameter $a/M$ is often used. 
The solutions of $\Delta = 0$ give the locations
of the inner and outer horizons,  $r_\pm=M\pm\sqrt{M^2-a^2}$, 
and the event horizon is located at $r=r_+$. 
In the Kerr geometry, the equation for the axion field is
\begin{multline}
-F{\varphi}_{,tt}
-2a(r^2+a^2-\Delta){\varphi}_{,t\phi}
+\frac{\Delta - a^2\sin^2\theta}{\sin^2\theta}\varphi_{,\phi\phi}
+\Delta\left(\varphi_{,\theta\theta}+\cot\theta\varphi_{,\theta}\right)
\\
+2r\Delta \varphi_{,r_*}+
(r^2+a^2)^2\varphi_{,r_*r_*}
-\Sigma\Delta \hat{U}^\prime(\varphi)=0,
\end{multline}
where
\begin{equation}
F:=(r^2+a^2)^2-\Delta a^2\sin^2\theta.
\label{Eq:definition_F}
\end{equation}
Here, we introduced
the tortoise coordinate by $dr_* = [(r^2+a^2)/\Delta] dr$, or
equivalently,
\begin{equation}
r_*=r+\frac{2M}{r_+-r_-}
\left(r_+\log|r-r_+|-r_-\log|r-r_-|\right).
\end{equation}
In the tortoise coordinate $r_*$, the horizon is located
at $r_*=-\infty$.

The scalar field in a Kerr spacetime has
conserved quantities.  
If a Killing vector $\xi^a$ is present in a spacetime, we can define
the conserved current $P_a=-T_{ab}\xi^b$ that satisfies $\nabla_aP^a=0$.
Here, $T_{ab}$ is the
energy-momentum tensor (in the unit $f_a=1$)
\begin{equation}
T_{ab}=\nabla_a\varphi\nabla_b\varphi
-\frac12 g_{ab}
\left(\nabla_c\varphi\nabla^c\varphi+2\hat{U}(\varphi)\right).
\end{equation}
Using this current $P_a$, the conserved quantity $C(t)\equiv C(0)$ 
can be introduced as
\begin{equation}
C(t):=\bar{C}(t) + \Delta C(t),
\end{equation}
with the quantity $\bar{C}(t)$ in the region 
$r_*^{\rm (in)}\le r_*\le r_*^{\rm (out)}$,
\begin{equation}
\bar{C}(t)=\int_{\Sigma_{t}} P_an^ad\Sigma, 
\label{Eq:barC}
\end{equation}
and the integrated flux toward the horizon at $r_*=r_*^{\rm (in)}$,
\begin{equation}
\Delta C(t)=\int_{r_*=r_*^{\rm (in)}} P_as^a d\sigma.
\label{Eq:Delta_C}
\end{equation}
Here, we have assumed the absence of outgoing flux 
at the outer boundary $r_*=r_*^{\rm (out)}$.
The integration of the first term $\bar{C}(t)$ is performed
on $t={\rm const.}$ slice $\Sigma_t$
in the range $r_*^{\rm (in)}\le r_*\le r_*^{\rm (out)}$
with the past-directed timelike unit normal $n^a$
and the volume element $d\Sigma$,
and it represents the conserved quantity contained in the 
region $r_*^{\rm (in)}\le r_*\le r_*^{\rm (out)}$.
The second term $\Delta C(t)$ is the integrated flux, where
the integration is performed on the hypersurface $r_{*}=r_*^{\rm (in)}$ 
from time zero to $t$ with the spacelike unit normal $s^a$
directing toward the horizon
and the surface element $d\sigma$. The value of $\Delta C(t)$
indicates the total quantity that has fallen into the BH from time
zero to $t$.
Since the Kerr spacetime possesses two Killing vectors, 
$\xi^a = (\partial_t)^a$ and $(\partial_\phi)^a$,
there exist two conserved quantities: the energy $E$ and 
the angular momentum $J$.

\subsection{Superradiant instability}
\label{Sec:superradiance}

Here, we briefly review the superradiant instability 
of a massive Klein-Gordon field 
around a Kerr BH. 
If the nonlinear terms are absent, the separation of
variables is available as follows. 
Setting $\varphi=2\mathrm{Re}[e^{-i\omega t}R(r)S_{\ell m}(\theta)e^{im\phi}]$,
the equations for $S_{\ell m}(\theta)$ and $R(r)$ become
\begin{equation}
\frac{1}{\sin\theta}
\frac{d}{d\theta}
\left(\sin\theta\frac{dS_{\ell m}}{d\theta}\right)
+\left[
-k^2a^2\cos^2\theta-\frac{m^2}{\sin^2\theta}+E_{\ell m}
\right]S_{\ell m}=0,
\label{Eq:angular_Teukolsky}
\end{equation} 
\begin{equation}
\frac{d}{dr}\left(\Delta
\frac{dR}{dr}\right)
+\left[
\frac{K^2}{\Delta}
-\lambda_{\ell m}
-\mu^2r^2
\right]R=0,
\label{Eq:radial_Teukolsky}
\end{equation}
where
\begin{equation}
K=(r^2+a^2)\omega-am,
\end{equation}
\begin{equation}
k^2=\mu^2-\omega^2,
\end{equation}
and
\begin{equation}
\lambda_{\ell m}=E_{\ell m}+a^2\omega^2-2am\omega.
\end{equation}
Here, $S_{\ell m}(\theta)e^{im\phi}$ is 
the spheroidal harmonics, which coincides with the
spherical harmonics in the case $k=0$. 
The angular quantum numbers, $\ell$ and $m$, are integers 
$\ell=0, 1, 2,...$ and $-\ell\le m\le \ell$. 
The eigenvalue $E_{\ell m}$ is $E_{\ell m}=\ell(\ell+1)$ in the case of $k=0$,
while in the case $k\neq 0$, 
it has to be evaluated numerically
by the methods of Refs.~\citen{Leaver:1985,Hughes:1999} 
or by the approximate formulas \cite{Brewer:1977,Seidel:1988,Berti:2005}.

From the equation \eqref{Eq:radial_Teukolsky}
for the radial function $R(r)$, 
the behavior of $R(r)$ at $r_*/M\gg 1$ is described as
$R\sim r^{-1}\exp(\pm kr)$.
If $\mathrm{Re}[\omega]< \mu$, the field is bounded by
gravitational interaction and does not escape to infinity. 
On the other hand, the behavior
of $R(r)$ in the neighborhood of the horizon $r_*/M\ll -1$ is 
$R\sim e^{\pm i\tilde{\omega}r_*}$, where
the plus and minus signs correspond to the outgoing and ingoing 
modes, respectively. Here, $\tilde{\omega}$ is defined as
$\tilde{\omega} = \omega-m\Omega_H$ 
with the angular velocity of the horizon $\Omega_H= a/(2Mr_+)$.

Here, let us focus attention to the energy 
$E$ of the Klein-Gordon field introduced
in Sec.~\ref{Sec:Field_Kerr}. 
Evaluating the energy density with respect to the
tortoise coordinate $r_*$ for the 
$t=\mathrm{const.}$ surface, we have 
$d\bar{E}/dr_*\simeq 2\omega\tilde{\omega}(r_+^2+a^2)$
in the neighborhood
of the horizon. Here, we have used the ingoing solution 
 $R\sim e^{-i\tilde{\omega}r_*}$ for $r_*/M\ll -1$.
On the other hand, 
the energy flux $F_E:=d(\Delta E)/dt$
toward the horizon can be evaluated as 
$F_E\simeq 2\omega\tilde{\omega}(r_+^2+a^2)$. 
Therefore, if waves satisfy the superradiant condition 
$0<\omega<m\Omega_H$ (i.e., $\tilde{\omega}<0$),
the negative energy distributes 
in the neighborhood of the horizon and it falls into the
BH ``at the speed of light'' in the coordinates $(t, r_*)$.

The negative energy of waves satisfying the superradiant condition
leads to an interesting phenomena. Suppose waves satisfying
the superradiant condition are incident to a rotating BH. 
A fraction of waves falls into the BH, and 
the rest is reflected back to infinity by the centrifugal
potential barrier of the BH. Since the
negative energy falls into the BH, the reflected waves
have greater energy than the initial ingoing waves
because of the energy conservation.
In other words, reflected waves  
get amplified. This 
is called superradiance. The superradiance was proposed
and analyzed for the massless Klein-Gordon field first 
by Zel'dovich \cite{Zel'dovich:1971,Zel'dovich:1972}.

Using superradiance, Press and Teukolsky \cite{Press:1972}
proposed a mechanism to cause an instability of fields
around a rotating BH, which is called the BH bomb. 
In this mechanism, a mirror is put around a BH.
Waves satisfying the superradiant condition are reflected back and
forth between the BH horizon and the mirror, and thus,
continue to get amplified. As a result, the amplitude
of waves exponentially grows in time. 
The mirror in the BH-bomb model seems to be artificial.
However, it was pointed out by Damour {\it et al.}~\cite{Damour:1976}
that if the field has non-vanishing mass, the reflected waves can
fall back to the BH because of the gravitational
force on the rest mass. In other words, if the field is in a quasibound state, 
$\mathrm{Re}[\omega] < \mu$, the superradiant instability occurs without 
putting a mirror. The instability of a massive Klein-Gordon field
around a Kerr BH was analytically studied by Detweiler 
\cite{Detweiler:1980}
and Zouros and Eardley \cite{Zouros:1979}.

Detweiler \cite{Detweiler:1980} 
analyzed the situation $\alpha_g:=M\mu \simeq M\omega \ll 1$. 
In this setup, the solution of the radial function $R(r)$
can be obtained by the matching method. After the solutions for the distant
region and the near-horizon region are obtained separately, 
they are matched to each other in an overlapping region.
After the matching, the solution for a distant region is same
as the wavefunction of the eigenstate of a 
hydrogen atom in quantum mechanics, since the
equation for the scalar field is same as the Schr\"odinger
equation for a hydrogen atom with the potential $e^2/r$
being replaced by $\alpha_g/r$. The result of the growth rate for
the $(\ell, m)=(1,1)$ mode is $\gamma M=(1/24)\alpha_g^9 (a/M)$.

%
\begin{figure}[tb]
\centering
{
\includegraphics[width=0.6\textwidth]{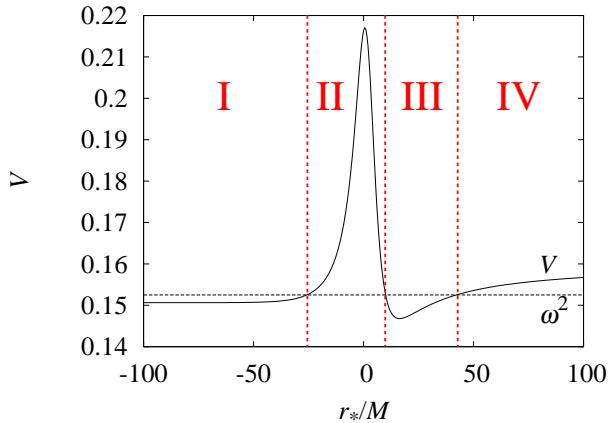}
}
\caption{
The potential $V(\omega,r_*)$ (in the unit $M=1$) in Eq.~\eqref{WKB-equation}
for a quasibound state of the Klein-Gordon field 
for situation $a/M=0.99$ and $\alpha_g:=M\mu=0.4$ (solid line).
The horizontal dotted line indicates the value of $\omega^2$.
Here, the imaginary part is ignored. 
There are four domains I, II, III, and IV, 
depending on the relation between
$V$ and $\omega^2$, and 
the quasibound state
is formed in region~III. Due to the tunneling effect, the waves
gradually fall into the region~I. Because the energy of waves
takes a negative value in region~I under the superradiant condition, 
the field in region~III is amplified.
}
\label{Fig:potential}
\end{figure}
%

Zouros and Eardley \cite{Zouros:1979} assumed $\alpha_g\gg 1$ and 
analyzed with the WKB approximation. Introducing
a function $u=\sqrt{r^2+a^2}R$, 
the radial mode equation is rewritten
as the Schr\"odinger-type equation:
\begin{equation}
\frac{d^2u}{dr_*^2}+
\left[\omega^2-V(\omega,r_*)\right]u
=0.
\label{WKB-equation}
\end{equation}
The potential for $a/M=0.99$ and $\alpha_g=0.4$ is shown in
Fig.~\ref{Fig:potential}. 
The potential $V$ asymptotes to $\mu^2$ 
from below for $r_*\to \infty$,
and this potential rise 
plays the role of the mirror. The quasibound state is formed
in the region~III, and because of the tunneling effect,  
the mode function gradually escape into the region~I
as ingoing waves. If these ingoing
waves toward the horizon satisfies the superradiant condition,
the energy of the quasibound state increases and the wavefunction
get amplified in the region~III. 
Their result shows that the growth rate
$M\gamma$ exponentially decreases as $\alpha_g$ is increased.

%
\begin{figure}[tb]
\centering
{
\includegraphics[width=0.4\textwidth]{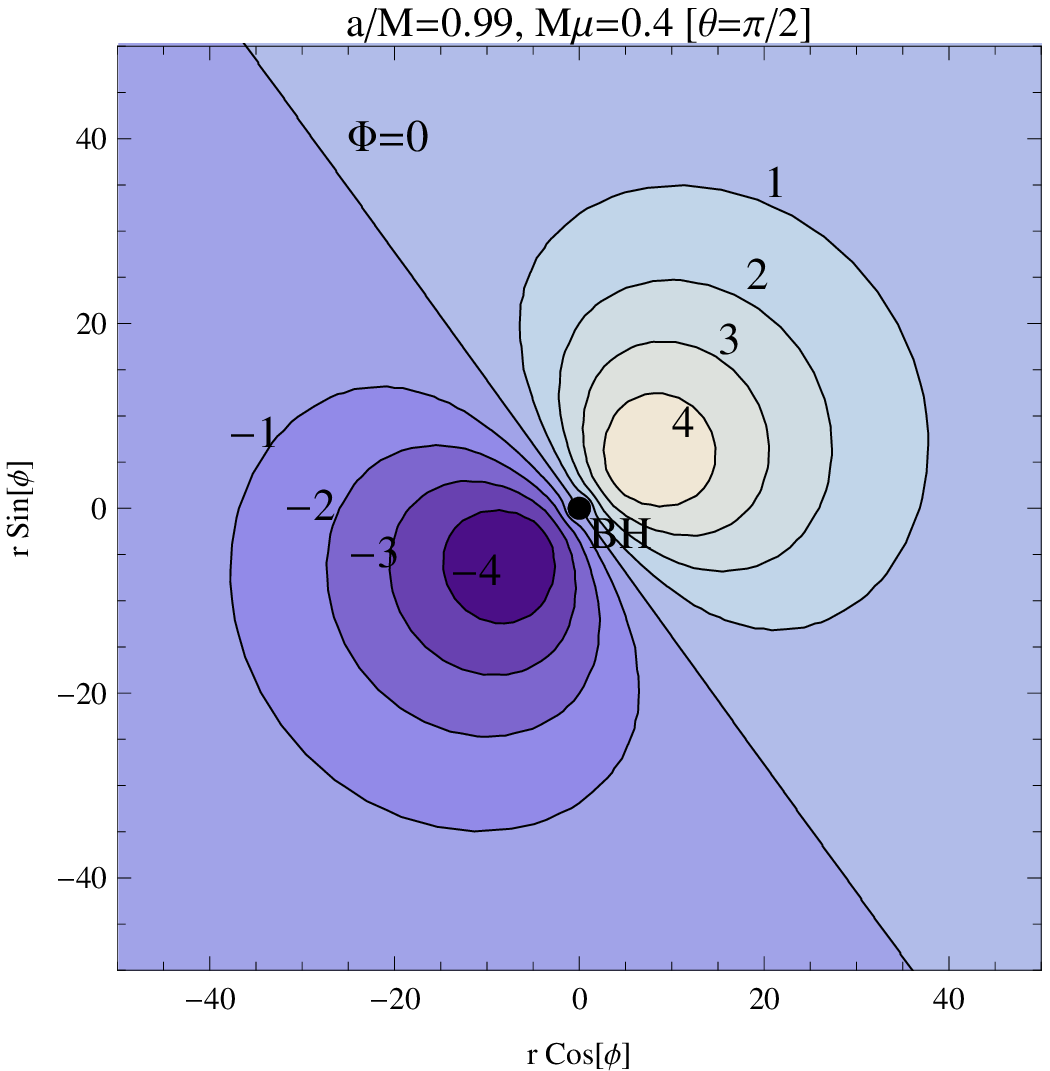}
\includegraphics[width=0.4\textwidth]{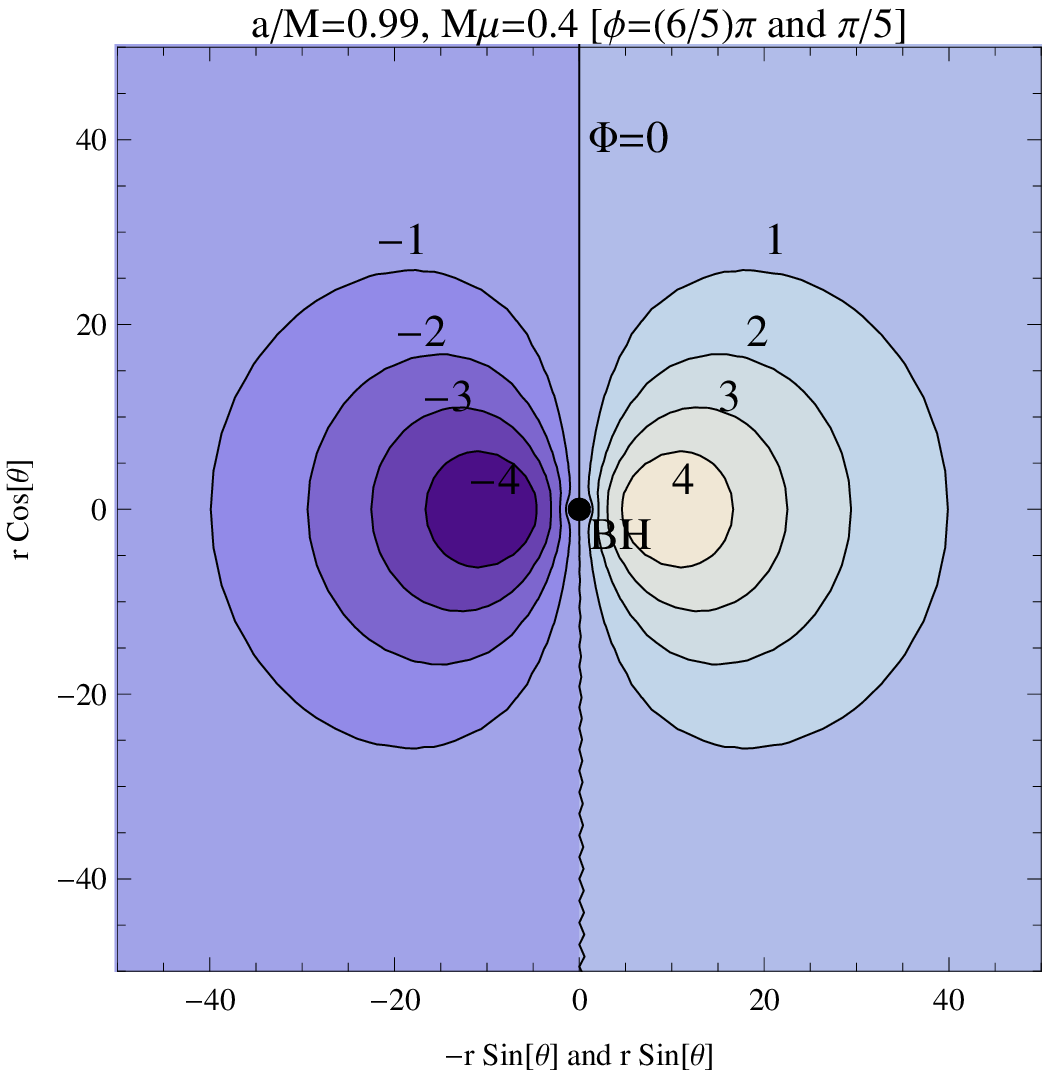}
}
\caption{A snapshot for the contours of the Klein-Gordon field $\varphi$
of the $(\ell,m)=(1,1)$ mode of the quasibound state in the case of
$a/M=0.99$ and $\alpha_g:=M\mu=0.4$ in the equatorial plane $\theta=\pi/2$
(left panel) and in the $(\rho, z)$-plane (right panel).
Here, $\rho:=r\sin\theta$ and $z:=r\cos\theta$,
and the $(\rho, z)$-plane is drawn for the azimuthal angle
$\phi = \pi/5$ and $(6/5)\pi$ so that the plane crosses the peak of the
field.
}
\label{Fig:quasibound_state}
\end{figure}
%

In the region where the largest growth rate of instability
is expected, $\alpha_g\sim 1$, numerical calculations are
required. These studies were done in 
Refs.~\citen{Furuhashi:2004,Strafuss:2004,Cardoso:2005,Dolan:2007}. 
The most detailed results
have been reported by Dolan \cite{Dolan:2007} by applying Leaver's
continued fraction method~\cite{Leaver:1985} to this problem.
The continued fraction method was originally developed to 
calculate the value of quasinormal frequencies numerically, and
it enables us to obtain highly accurate values of $\omega$
for the quasibound state as well. The result is shown 
in Figs.~6 and 7 of Ref.~\citen{Dolan:2007}.
The largest growth rate is realized for 
$(\ell, m)=(1,1)$, $a/M\simeq 1$, 
and $\alpha_g\simeq 0.4$, and its value is $\gamma/\mu \simeq 3\times 10^{-7}$.
In Ref.~\citen{Kodama:2011}, we also developed a code to calculate 
$\omega$ of the quasibound state and reproduced Dolan's result.
As an example, the configuration of the field $\varphi$ of the 
$(\ell,m)=(1,1)$ mode 
in the equatorial plane and in the $(\rho, z)$-plane
(where $\rho:=r\sin\theta$ and $z:=r\cos\theta$)
are shown as contour plots 
in the left and right panels of Fig.~\ref{Fig:quasibound_state},
respectively, for $a/M=0.99$ and $\alpha_g=0.4$.

%
%
\section{Numerical method and code}
\label{Sec:Techniques}

In this section, we explain the
technical part of our study. The formulation
for solving the axion field around a Kerr BH
is explained in Sec.~\ref{Sec:method}, and the numerical
techniques and code tests are summarized in 
Sec.~\ref{Sec:code}.

\subsection{Numerical method}
\label{Sec:method}

The most important point in simulations of fields
in the Kerr background spacetime is to realize the
sufficient stability. 
We found that 
if a simulation is performed
in the Boyer-Lindquist coordinates with central difference method,
a numerical instability immediately develops to crash the simulation.
This is because the lines with constant spatial coordinates 
are spacelike in the ergoregion in the Boyer-Lindquist coordinates, 
and therefore, the frame is propagating superluminally. 
Although this problem may be avoided by adopting the upwind 
difference method, we have chosen another method
with which greater stability is expected.
This method is explained in Secs.~\ref{Sec:ZAMO} and \ref{Sec:pullback}.
We also explain the boundary condition and how to 
regularize the equation at the poles in Secs.~\ref{Sec:boundary}
and \ref{Sec:pole}, respectively.

\subsubsection{ZAMO coordinates}
\label{Sec:ZAMO}

In our method, we realize the numerical stability 
by adopting the coordinates
associated with the zero-angular-momentum observers (ZAMOs).
The ZAMOs are observers such that they stay at fixed 
$r$ and $\theta$, but move in the $\phi$ direction 
so that their angular momenta are kept to be zero. 
Their four-velocity is 
given by $u^a=\nabla^at/\sqrt{-\nabla^bt\nabla_bt}$,
which is timelike everywhere, 
and they rotate with the angular velocity
\begin{equation}
\Omega (r,\theta)
=\frac{2Mar}{F},
\end{equation}
where $F$ is defined in Eq.~\eqref{Eq:definition_F}. 
Using this angular velocity, we introduce the
new coordinates $(\tilde{t}, \tilde{\phi}, \tilde{r}, \tilde{\theta})$
as
\begin{equation}
\tilde{t} = t,\quad 
\tilde{\phi}  =  \phi - \Omega(r,\theta)t,\quad 
\tilde{r} = r,\quad 
\tilde{\theta} = \theta.
\end{equation}
The basis vector of the new time coordinate $\tilde{t}$
is parallel to $u^a$, and therefore, it is
timelike everywhere. 
We call these coordinates the ZAMO coordinates.
The equation for the axion field
reads
\begin{multline}
-F\varphi_{,\tilde{t}\tilde{t}}
+\left[\frac{\Sigma^2\Delta}{F\sin^2\tilde{\theta}}
+\tilde{t}^2{\Delta}\left(\Delta \Omega_{,\tilde{r}}^2
+\Omega_{,\tilde{\theta}}^2\right)
\right]
\varphi_{,\tilde{\phi}\tilde{\phi}}
+{(\tilde{r}^2+a^2)^2}\varphi_{,\tilde{r}_*\tilde{r}_*}
+2\tilde{r}\Delta\varphi_{,\tilde{r}_*}
\\
+{\Delta}\left(
\varphi_{,\tilde{\theta}\tilde{\theta}}+\cot\tilde{\theta}\varphi_{,\tilde{\theta}}
\right)
-2\tilde{t}
{\Delta}
\left[(\tilde{r}^2+a^2)\Omega_{,\tilde{r}}\varphi_{,\tilde{r}_*\tilde{\phi}}
+\Omega_{,\tilde{\theta}}\varphi_{,\tilde{\theta}\tilde{\phi}}
\right]
\\
-\tilde{t}{\Delta}
\left[
(\Delta\Omega_{,\tilde{r}})_{,\tilde{r}}
+(\Omega_{,\tilde{\theta}\tilde{\theta}}
+\cot\tilde{\theta}\Omega_{,\tilde{\theta}})
\right]\varphi_{,\tilde{\phi}}
-{\Sigma\Delta}\hat{U}^\prime(\varphi)=0,
\label{Eq:ZAMO-coordinate}
\end{multline}
in the ZAMO coordinates.

\subsubsection{Pullback of coordinates}
\label{Sec:pullback}

Because the
angular velocity $\Omega$ of a ZAMO becomes larger as it is closer
to the horizon,
the ZAMO coordinates 
become distorted in time evolution.
This is the shortcoming of the ZAMO coordinates
because  
if the coordinates are distorted, the numerical error
grows large, and also, the physical interpretation
of the numerical results becomes difficult.
We solve this problem by ``pulling back'' the ZAMO
coordinates. Namely, when the ZAMO coordinates become 
distorted to some extent, we introduce new ZAMO coordinates
which are not distorted at that time (i.e., the new
ZAMO coordinates agree instantaneously with the Boyer-Lindquist
coordinates), and continue 
time evolution with the new coordinates. Iterating these
processes, longterm evolution becomes feasible.
Specifically, for $nT_P\le t\le (n+1)T_P$, we adopt the $n$-th ZAMO
coordinates $(\tilde{t}^{(n)}, \tilde{\phi}^{(n)}, \tilde{r}^{(n)},
\tilde{\theta}^{(n)})$ by
\begin{equation}
\tilde{t}^{(n)} = t,\quad
\tilde{\phi}^{(n)}  =  \phi - \Omega(r,\theta)(t-nT_P),\quad
\tilde{r}^{(n)} = r,\quad
\tilde{\theta}^{(n)} = \theta.
\end{equation}
The numerical data of $\Phi$ and $\partial\Phi/\partial\tilde{t}$ in the
new coordinates are 
generated by interpolation. In our numerical calculations,
we adopt $T_P=M/4$. If we list up the data of $\Phi$ at time
$t=nT_P$ with $n=0,1,2,...$, they can be regarded as
the data in the Boyer-Lindquist coordinates.

\subsubsection{Boundary conditions}
\label{Sec:boundary}

Since a simulation has to be performed 
in a finite coordinate region, the coordinate 
range of $r_*$ is taken as
$r_*^{\rm (in)}\le r_*\le r_*^{\rm (out)}$.
Here, we discuss how to impose the inner and outer boundary conditions
at $r_*=r_*^{\rm (in)}$ and $r_*^{\rm (out)}$,
respectively.

For a sufficiently small $r_*^{\rm (in)}$, 
we have $\Delta \simeq 0$ at $r=r_*^{\rm (in)}$. Then, 
the equation for $\varphi$ in the
ZAMO coordinates, 
Eq.~\eqref{Eq:ZAMO-coordinate}, becomes
\begin{equation}
-\varphi_{,\tilde{t}\tilde{t}}+ \varphi_{\tilde{r}_*\tilde{r}_*}\simeq 0.
\end{equation}
Therefore, the in- and out- going modes are clearly separated,
and we can impose the purely ingoing boundary condition
in the standard manner. Typically, we adopt 
$r_*^{\rm (in)}/M=-200$.

At $r=r_*^{\rm (out)}$, we adopt the fixed boundary condition,
$\varphi=0$. When the axion field is in a bound state,
this boundary condition gives a good approximation.
If outgoing waves are generated, the outer boundary
becomes reflective, which is quite artificial.
In such a case, we avoid the problem by adopting
sufficiently large $r_*^{\rm (out)}$. Typically, 
the outer boundary is located between  
$r_*^{\rm (out)}/M=200$ and $1000$ depending on the situation.

\subsubsection{Regularization at poles}
\label{Sec:pole}

Since the two poles $\tilde{\theta} = 0$ and $\pi$
are coordinate singularities, regularization 
of the equation is required at the poles.
For this purpose, we introduce new coordinates
$(x,y)$ by
\begin{equation}
x=\tilde{\theta}\cos\tilde{\phi}, \quad 
y=\tilde{\theta}\sin\tilde{\phi},
\end{equation}
in the neighborhood of each pole. 
Rewriting Eq.~\eqref{Eq:ZAMO-coordinate}
with these coordinates and taking the limit $\tilde{\theta}\to 0$ or $\pi$,
we obtain
\begin{equation}
-F{\varphi}_{,\tilde{t}\tilde{t}} 
+{(\tilde{r}^2+a^2)^2}\varphi_{,\tilde{r}_*\tilde{r}_*}
+{\Delta}
\left[
\varphi_{,xx}+\varphi_{,yy}
+2\tilde{r}\Delta \varphi_{,\tilde{r}_*}
-\Sigma \hat{U}^\prime(\varphi)
\right]=0.
\end{equation}
Here, $\varphi_{,xx}$ can be evaluated by 
the data at the grids on $\tilde{\phi}=0$ and $\pi$,
and $\varphi_{,yy}$ by the data at the grids on $\tilde{\phi}=\pi/2$
and $(3/2)\pi$. Therefore, the data of grids at 
the poles can be evolved toward
the next time step with this equation.

\subsection{Code and code checks}
\label{Sec:code}

Our code is a three-dimensional (3D) code of 
the ZAMO coordinates $(\tilde{r}_*,\tilde{\theta},\tilde{\phi})$.
The sixth-order finite differencing method is used in
spatial directions, and time evolution is proceeded
with the fourth-order Runge-Kutta method.
Typically, we used the grid size $\Delta r_*/M=0.5$ and
$\Delta\theta=\Delta\phi=\pi/30$. When the spherical-polar
coordinates are used, the Courant condition 
for the time step becomes severe and it has to be chosen 
so that $\Delta t\lesssim \mathrm{min}[(F/\Sigma\Delta^{1/2})_{\theta=0}
\Delta\theta\Delta\phi]$ from Eq.~\eqref{Eq:ZAMO-coordinate}. Here we adopt
the value of the time step as
$\Delta t=(3/2\pi)\Delta\theta\Delta r_*$ in order to realize the
sufficient stability of our simulations.
In doing the ``pullback'' of the ZAMO 
coordinates addressed in Sec.~\ref{Sec:pullback},
the interpolation of data is necessary, and we applied the
seventh-order Lagrange interpolation.

In order to validate the code, we have to
perform test simulations. The code checks
have been done in the three following manners,
as explained one by one below.

\subsubsection{Comparison with semianalytic solution}
\label{Sec:comparison}

The first check is to simulate time evolution of 
the quasibound state of the linear Klein-Gordon equation
and compare the numerical data 
with the semianalytic solution. 
Here, we choose the BH with the rotation parameter $a/M=0.99$
and the Klein-Gordon field of mass $\mu=0.4/M$. 
The semianalytic solution 
$\varphi=e^{-i\omega t}R(r)S(\theta)e^{im\phi}$
can be obtained by using the approximate formula
for $S_{\ell m}(\theta)$ \cite{Seidel:1988} 
and numerically calculating $\omega$
and $R(r)$ using the continued fraction method \cite{Leaver:1985,Dolan:2007}.
The time evolution was performed up to $t=100M$,
and the numerical data were confirmed to agree well
with the semianalytic solution.

The imaginary part $\gamma$ of frequency $\omega=\omega_0+i\gamma$
of the semianalytic solution gives the correct growth
rate of the superradiant instability.
In the present setup, it is calculated as $\gamma/\mu
\simeq 3.311\times 10^{-7}$ by the continued fraction method. 
In order to check to what extent the superradiant instability is
correctly realized in our numerical simulation, we calculated the energy 
$\bar{E}(t)$ in the region $r_*^{\rm (in)}\le r_*\le r_*^{\rm (out)}$
[see Eq.~\eqref{Eq:barC} in Sec.~\ref{Sec:Field_Kerr} for the definition 
of $\bar{E}(t)$], 
and evaluated 
$\gamma=(d\bar{E}/dt)/2\bar{E}
\simeq [\bar{E}(t_f)-\bar{E}(0)]/[2t_f\bar{E}(0)]$
with $t_f=100M$. 
The numerical result performed in  
the grid sizes mentioned above 
is $\gamma/\mu \simeq 3.255\times 10^{-7}$:
The deviation from the value of the semianalytic 
solution is about 1.7\%. 
Therefore, our code has the ability to describe the energy extraction
by the superradiant instability fairly accurately.

\subsubsection{Convergence with respect to grid size}

%
\begin{figure}[tb]
\centering
{
\includegraphics[width=0.6\textwidth]{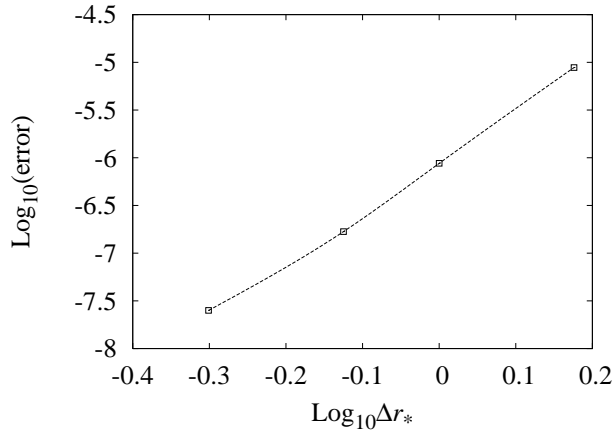}
}
\caption{The relation between the grid size $\Delta r_*$ (with 
unit $M=1$) and the numerical error evaluated at $t=12.5M$.
The error decreases as $\Delta r_*$ is increased,
and the slope of the curve is $\simeq 5$. This reflects
our combined fourth- and sixth-order scheme.
}
\label{Fig:Convergence}
\end{figure}
%

One of the standard tests of numerical simulations
is to check whether the numerical solution converges as the grid size
is made smaller. For this purpose, we adopt the numerical solution of
$\Delta r_*/M = 1/6$ as the reference solution, and evaluated
the deviation of the numerical solutions
with several grid sizes. Here, we 
adopted $\varphi(0)= \exp[(r_*/30)^2]\sin\theta\cos\phi$ and $\dot\varphi(0)=0$
as the initial condition and evolved the data until $t/M=12.5$
for the parameters $\alpha_g:=M\mu=0.4$ and $a/M=0.99$. 
Figure~\ref{Fig:Convergence} shows the relation between 
$\log_{10}\Delta r_*$ and  $\log_{10}({\rm error})$.
Since we use 
the sixth- and fourth-order schemes in the space and time directions,
the curve is expected to 
have slope between four and six.
Actually, the slope is $\sim 5$ in this figure. 
This result reflects the adopted scheme, and supports the validity
of our code.

\subsubsection{Conserved quantities}
\label{Sec:Conserved_quantities}

As discussed in Sec.~\ref{Sec:superradiance},
we have the two conserved quantities, the energy $E$ and
the angular momentum $J$. In actual simulations, 
these quantities slightly change in time because of
numerical error. Therefore, 
the deviations of the values $E(t)/E(0)$
and $J(t)/J(0)$ from unity give indicators
for the accumulated numerical errors.

%
\begin{figure}[tb]
\centering
{
\includegraphics[width=0.45\textwidth]{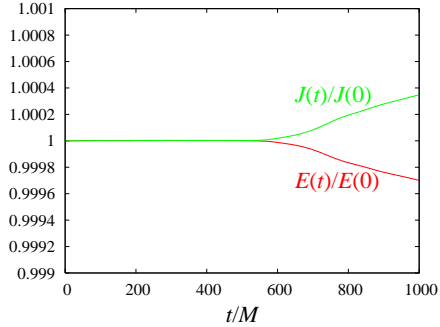}
}
\caption{The values of total energy and angular momentum 
normalized by the initial values, $E(t)/E(0)$ and $J(t)/J(0)$,
as functions of time (the solid line and the dashed line,
respectively). Deviation from unity indicates the 
amount of numerical error. The error is less than $0.04\%$ 
at $t/M=1000$.}
\label{Fig:EJ-time}
\end{figure}
%

Figure~\ref{Fig:EJ-time} shows the values of
$E(t)/E(0)$ and $J(t)/J(0)$ as functions of time $t/M$.
Here, we show the results for the simulation
of axion mass $\alpha_g=0.4$ around a BH
with $a/M=0.99$
for initial amplitude $\varphi_{\rm peak}(0)=0.7$ [i.e., simulation
(B) of Sec.~\ref{Sec:simulation_B}]. The deviation from unity
is negligible for $t/M\lesssim 500$. For $t\gtrsim 500$,
the deviations linearly increase. This is because
the ``bosenova'' happens and some part of the axion field distributes
at a distant place. As a result, small error in the field value
results in large errors of $E(t)$ and $J(t)$
because large volume element is multiplied there.
Nevertheless the deviations from unity are less than $0.04\%$ at $t/M=1000$
for both $E(t)/E(0)$ and $J(t)/J(0)$.

As found above, we checked the validity of our code
in three ways, and therefore, 
we can trust the results of our longterm simulations.

%
%
\section{Numerical results}
\label{Sec:results}

Now we present the numerical results. 
In Sec.~\ref{Sec:setup}, we describe
the setup of the system and the initial conditions.
In Sec.~\ref{Sec:two_simulations}, 
we show the results of typical two simulations
[referred as simulations (A) and (B)],
for which the effect of nonlinearlity is weak and strong.
respectively. This helps us to understand how nonlinearlity
works in this system. Then, in Sec.~\ref{Sec:bosenova},
we discuss what actually happens in the final stage of
the superradiant instability, taking special attention
to whether the bosenova happens or not.

\subsection{Setup}
\label{Sec:setup}

In order to study the nonlinear self-interaction
of an axion field, we numerically solve the
sine-Gordon equation $\Box\varphi-\mu^2\sin\varphi=0$
in a Kerr spacetime. 
For simplicity, we consider
an axion cloud with mass $\alpha_g:=M\mu=0.4$ 
around a Kerr BH with the rotational parameter $a/M=0.99$.  
As the initial
condition, we adopt the quasibound state solution to 
the linear Klein-Gordon field corresponding to the $(\ell,m)=(1,1)$ mode. 
Namely, the configuration shown in Fig.~\ref{Fig:quasibound_state}
are used (but changing the amplitude of the oscillation). 
If nonlinear terms are absent (i.e., in the case of
the Klein-Gordon field), 
the frequency is 
$\omega =\omega_0+i\gamma$, where
the real part is $\omega_0M\simeq 0.39$
and the imaginary part 
(i.e., the growth rate by 
the superradiant instability)  
is $\gamma M \simeq 1.32\times 10^{-7}$,
which is the approximately largest possible growth rate.
These are natural setups, because
we consider the situation where 
the axion fields have grown due to the superradiant instability
and, at least for small $|\varphi|$, such situations
should be well approximated by the quasibound state of
the Klein-Gordon field.  As typical examples, 
we present the results of two simulations 
with different initial amplitude
[simulations (A) and (B) in Secs.~\ref{Sec:simulation_A} and
\ref{Sec:simulation_B}, respectively, see Table~\ref{Table:simulations}]. 
In Sec.~\ref{Sec:bosenova}, 
we discuss what actually happens as a result
of superradiant instability by performing supplementary
simulations starting with 
initial conditions that is expected to be more natural
compared to those of the simulations (A) and (B).

\subsection{Typical two simulations}
\label{Sec:two_simulations}

%
\begin{table}[tb]
\caption{Performed two simulations, (A) and (B), 
in Sec.~\ref{Sec:two_simulations}.
``KG bound state'' means that the initial condition is adopted
as the quasibound state of Klein-Gordon field of $(\ell,m)=(1,1)$ mode,
and $\varphi_{\rm peak}(0)$ indicates the initial amplitude. }
\begin{tabular}{c|cccc}
\hline\hline
Simulations & Initial condition & ${E}/[(f_a/M_p)^2M]$ & nonlinearlity\\
  \hline
(A) & KG bound state, $\varphi_{\rm peak}^{\rm (A)}(0) = 0.60$ & 1430 & weak\\
(B) & KG bound state, $\varphi_{\rm peak}^{\rm (B)}(0) = 0.70$ & 1862 & strong\\
\hline\hline
\end{tabular}
\label{Table:simulations}
\end{table}
%

Now, we present the results of the simulations (A) and (B). 

\subsubsection{Simulation (A): A weakly nonlinear case}
\label{Sec:simulation_A}

In the simulation (A), we choose the initial amplitude
to be $\varphi_{\rm peak}(0)=0.6$. The effect of the nonlinearlity
can be evaluated by 
$\Delta_{\rm NL}:=(\varphi-\sin\varphi)/\varphi \simeq \varphi^2/6$, 
and for this setup, $\Delta_{\rm NL}=0.06$ at the peak. Therefore,
the nonlinear effects are weakly important for this
situation.

%
\begin{figure}[tb]
\centering
{
\includegraphics[width=0.6\textwidth]{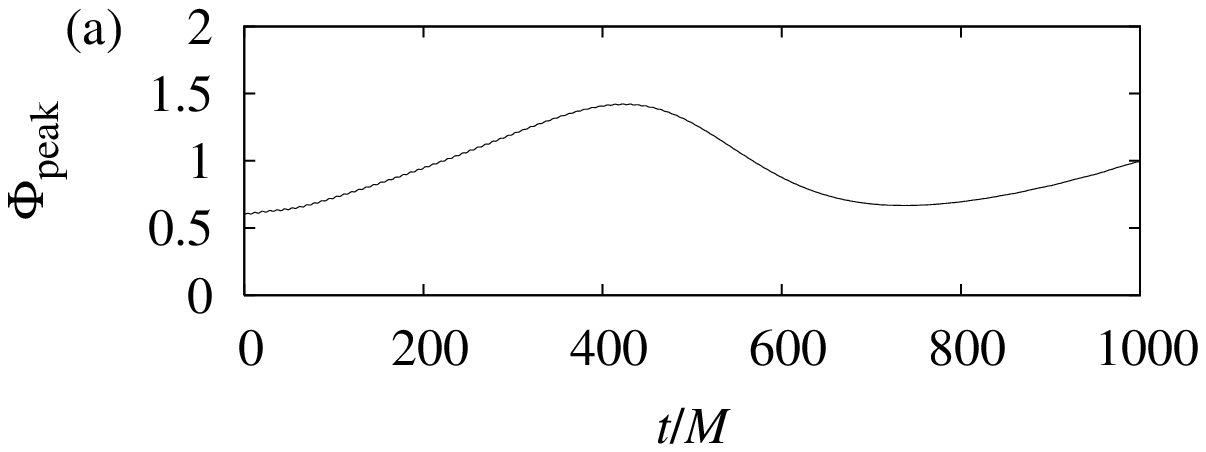}
\includegraphics[width=0.6\textwidth]{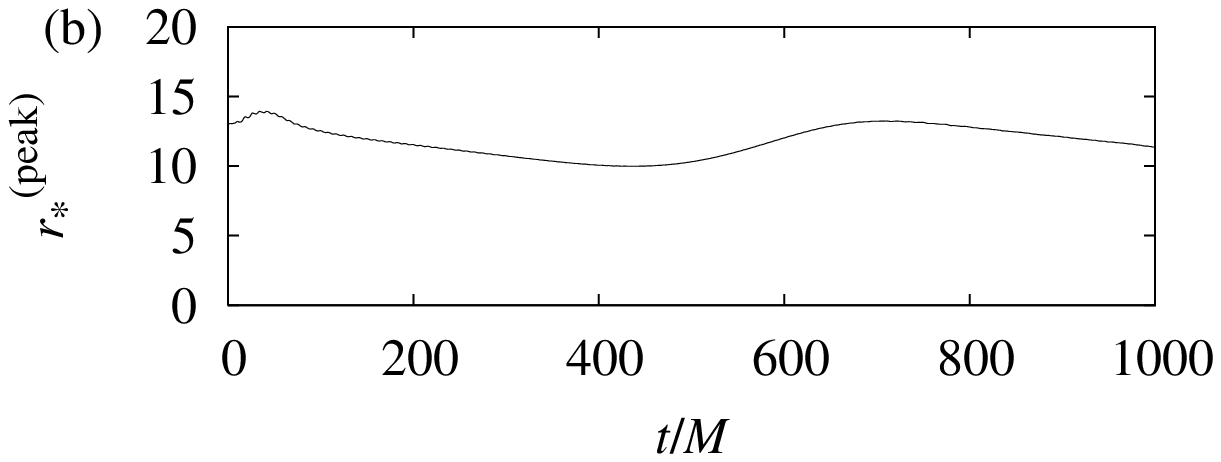}
}
\caption{The peak value $\varphi_{\rm peak}$ of the field 
$\varphi$ (upper panel) and its location $r_*^{\rm (peak)}$ with respect 
to the tortoise coordinate (lower panel) as functions of time
observed in simulation (A)
[i.e., $\varphi_{\rm peak}(0)=0.6$]. 
The peak location moves back and forth periodically. 
When the peak location becomes
close to the horizon, the value of $\varphi_{\rm peak}$
becomes larger. 
}
\label{Fig:amplitude_0.6}
\end{figure}
%

The upper panel 
of Fig.~\ref{Fig:amplitude_0.6} shows the value of the field
at the peak, $\varphi_{\rm peak}:=\mathrm{sup}[\varphi]$, 
and the lower panel shows
the position $r_*^{\rm (peak)}$ of the peak 
with respect to the tortoise coordinate $r_*$
as functions of $t/M$. 
The value of $\varphi_{\rm peak}$ oscillates with the period
of about $700M$. The position of the peak also moves back and forth,
and $\varphi_{\rm peak}$ increases when $r_*^{\rm (peak)}$
decreases, i.e., when the peak location approaches the horizon.
Therefore, the change in the amplitude is mainly caused not by
the superradiant instability but by the change of the peak position.
Namely, when the peak position approaches the horizon
due to nonlinear interaction, the field gets compacted
in a small region around the BH, and therefore, the field is amplified
because of the energy conservation. 

%
\begin{figure}[tb]
\centering
{
\includegraphics[width=0.9\textwidth]{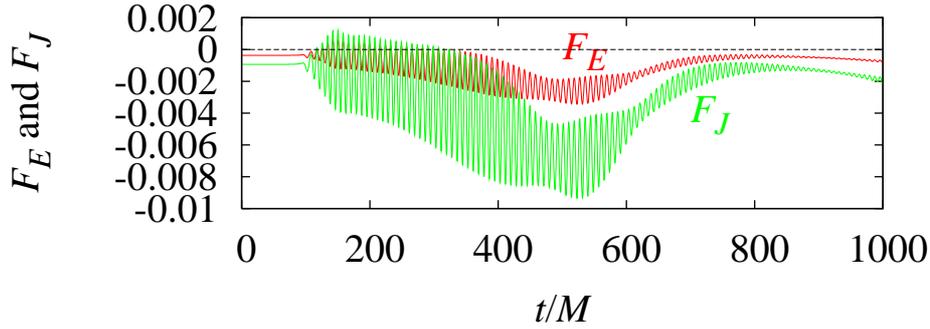}
}
\caption{Fluxes $F_E$ and $F_J$ of energy and angular momentum,
respectively, toward the horizon observed in simulation (A)
[i.e., $\varphi_{\rm peak}(0)=0.6$]. $F_E$ and $F_J$ 
are negative except for
very short periods. Therefore, the energy and angular momentum
are extracted from the BH. 
The nonlinear effect makes their values larger.
}
\label{Fig:EJ-flux-amp0.6}
\end{figure}
%

Figure~\ref{Fig:EJ-flux-amp0.6} shows 
the energy flux $F_E:=d(\Delta E)/dt$
and the angular momentum flux $F_J:=d(\Delta J)/dt$
toward the horizon evaluated at $r_*=-100M$,
where $\Delta E$ and $\Delta J$ are integrated fluxes 
defined in Eq.~\eqref{Eq:Delta_C}. 
Initially, both $F_E$ and $F_J$ are negative, reflecting
the fact that we have chosen the $(\ell,m)=(1,1)$ mode of the superradiant
bound state as the initial condition. 
The nonlinear effect appears at $t/M\gtrsim 100$.
In that period, both $F_E$ and $F_J$ oscillate 
rapidly, and their mean values
are negative. The absolute values of
$F_E$ and $F_J$ become larger around $t=500M$. 
The primary nonlinear effects are the 
following two. The first effect is that it enhances the amplitude
of the waves of the $(\ell,m)=(1,1)$ mode that fall into the BH
scarcely changing
the real part of the frequency $\omega_0$. 
The second effect is that it generates waves of the 
$(\ell,m)=(1,-1)$ mode
with frequency 
$\omega_{\rm NL}$
which is approximately same as that of waves of the $(\ell,m)=(1,1)$ mode,
$\omega_{\rm NL}\approx \omega_0$.
Although waves of the $(\ell, m)=(1,-1)$ mode 
generate the positive energy flux to the horizon,
it is very small in this case. 
Therefore, the first effect is much stronger
than the second effect, and 
the rates of the extraction of energy 
and angular momentum are enhanced.
By the interference of the two modes, 
the small oscillation of $F_E$ and $F_J$ 
appear with frequency $\approx 2\omega_0$. 
The generation of waves of the $(\ell, m)=(1,-1)$ mode 
becomes more significant
in the strongly nonlinear case [the simulation (B)]
as explained later. Although the generation of the $(\ell, m)=(1,-1)$ mode 
may seem strange because naively we expect the nonlinearlity
to produce modes in proportion to $e^{\pm i n(m\phi-\omega t)}$ 
with integer $n$ from $\mathrm{Re}[e^{i(m\phi-\omega t)}]$, 
the Green's function analysis in 
Appendix~\ref{Sec:Green} supports 
it (see also Sec.~\ref{Sec:simulation_B}).

%
\begin{figure}[tb]
\centering
{
\includegraphics[width=0.45\textwidth]{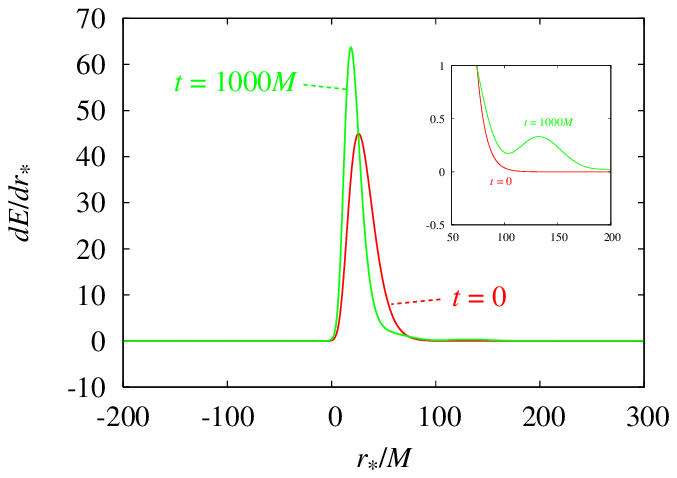}
\includegraphics[width=0.45\textwidth]{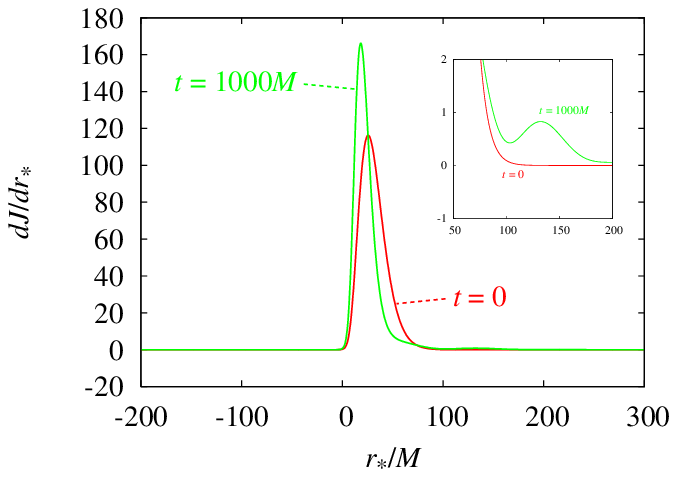}
}
\caption{The energy density $dE/dr_*$ (left) and the angular momentum 
density $dJ/dr_*$ (right) with respect to the tortoise coordinate $r_*$
at time $t/M=0$ and $1000$ for simulation (A) 
[i.e., $\varphi_{\rm peak}(0)=0.6$]. 
}
\label{Fig:dEdr-dJdr-amp0.6}
\end{figure}
%

The left and right panels 
of Fig.~\ref{Fig:dEdr-dJdr-amp0.6} show the energy density
$dE/dr_*$ and the angular momentum density $dJ/dr_*$
with respect to the tortoise coordinate $r_*$, respectively, 
at $t/M=0$ and $1000$.  
There are two peaks for the curve of $t/M=1000$ in each panel, 
the first peak near the horizon
and the second peak at $r_*/M\simeq 140$ (as
can be seen in the inset). 
The locations of the first peaks of energy and angular momentum densities
are shifted to small $r_*$ values
compared to $t=0$. This means that most of the energy 
gets squeezed into a small region 
close to the horizon 
because of the nonlinear attractive self-interaction. 
Another effect of the nonlinearlity is that it transports 
a small fraction 
of energy and angular momentum to a region far from the BH,
making the small second peak as seen in the inset of each panel.

\subsubsection{Simulation (B): A strongly nonlinear case}
\label{Sec:simulation_B}

In the simulation (B), we choose the initial amplitude
to be $\varphi_{\rm peak}(0)=0.7$. The 
parameter $\Delta_{\rm NL}\simeq \varphi^2/6$ for the effect of 
the nonlinearlity
is $\Delta_{\rm NL}\simeq 0.082$ at the peak for this setup,
and therefore,
the nonlinear effect is larger compared to the simulations (A). 
The nonlinearlity in this
situation is strong enough
for causing the bosenova collapse.

%
\begin{figure}[tb]
\centering
{
\includegraphics[width=0.4\textwidth]{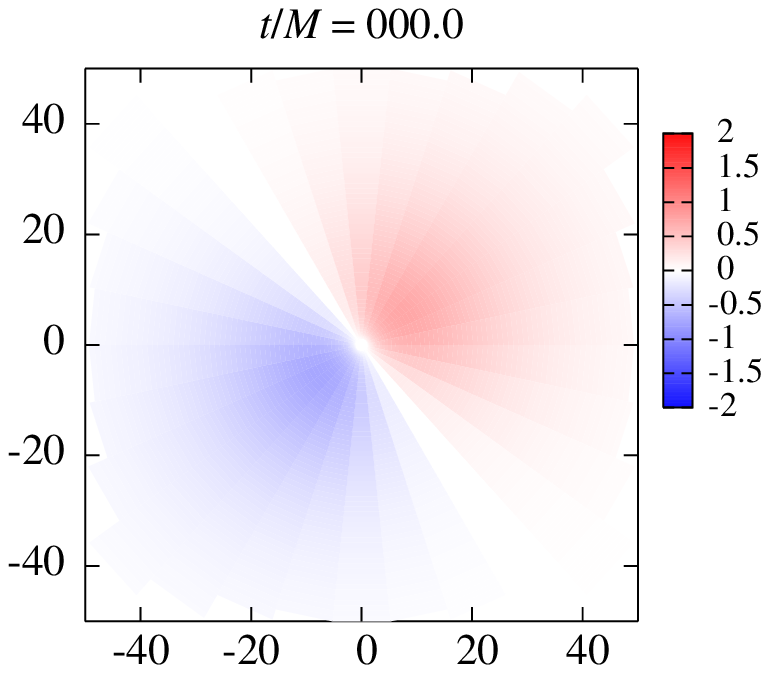}
\includegraphics[width=0.4\textwidth]{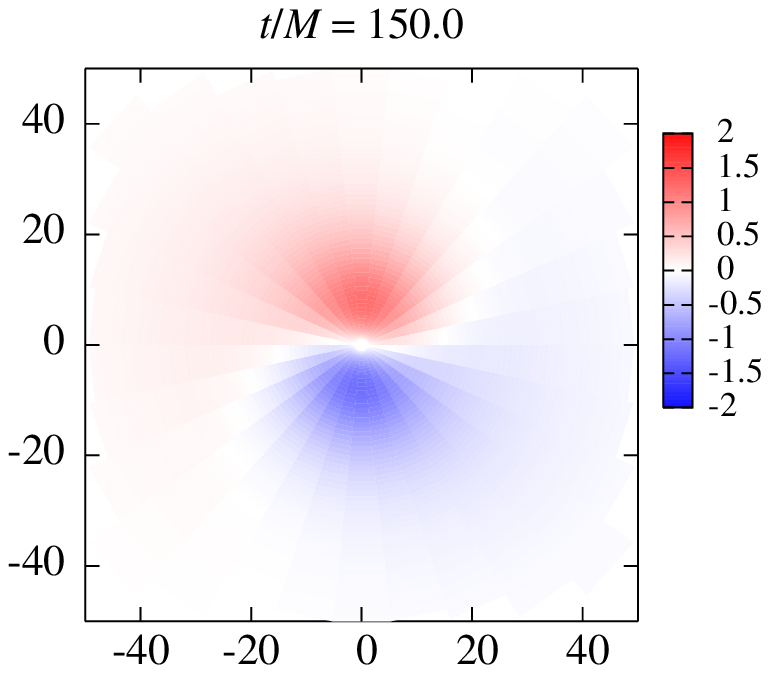}
\\
\includegraphics[width=0.4\textwidth]{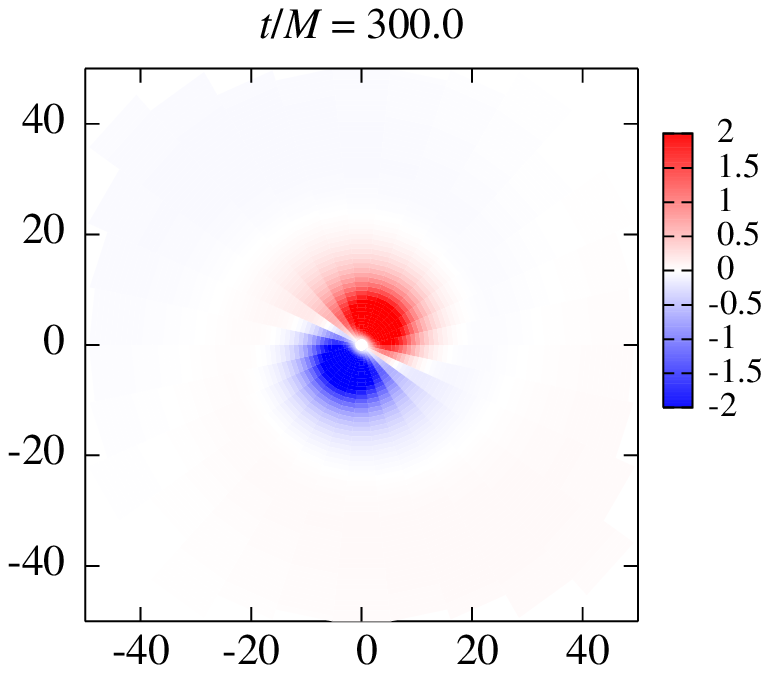}
\includegraphics[width=0.4\textwidth]{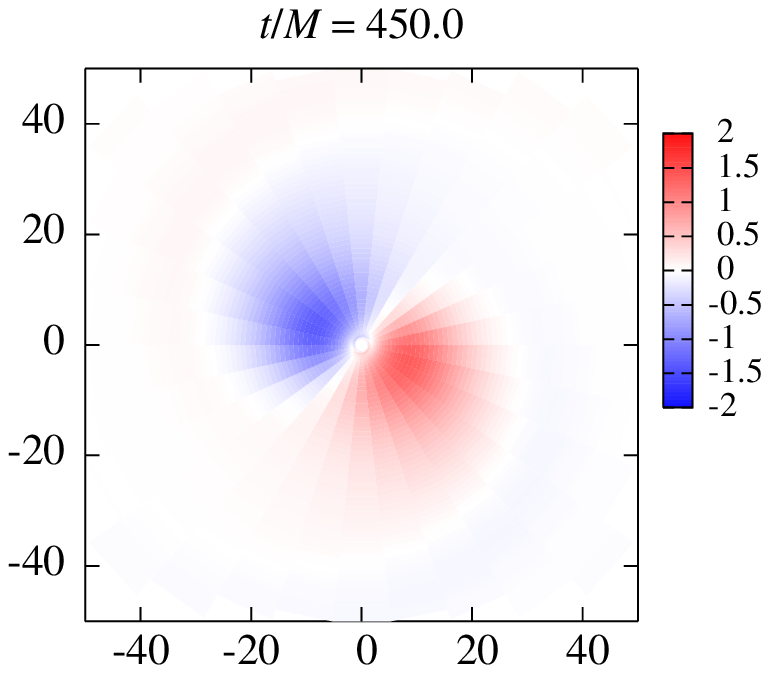}
}
\caption{Snapshots of density plot of the axion field $\varphi$ 
in the equatorial $(r\cos\phi, r\sin\phi)$-plane 
($\theta=\pi/2$) at $t/M=0$, $150$, $300$, and $450$. The 
axion cloud is rotating counterclockwise.
}
\label{Fig:density-snapshots}
\end{figure}
%

First, we would like to present some snapshots.
Figure~\ref{Fig:density-snapshots} 
shows the density plots of the axion field $\varphi$ 
in the equatorial $(r\cos\phi, r\sin\phi)$-plane 
($\theta=\pi/2$) at $t/M=0$, $150$, $300$, and $450$.
The initial condition $t=0$ is the bound state of the 
Klein-Gordon field.
In the time evolution, the axion cloud rotates counterclockwise
and gradually becomes closer to the BH ($t=150$).
At $t=300$, the axion
cloud is highly concentrated in a small region around the BH,
and this is when the bosenova begins to happen. 
During the bosenova, the shape of the
axion cloud becomes distorted and part of the cloud is scattered
to the distant region and into the BH ($t=450$).

%
\begin{figure}[tb]
\centering
{
\includegraphics[width=0.6\textwidth]{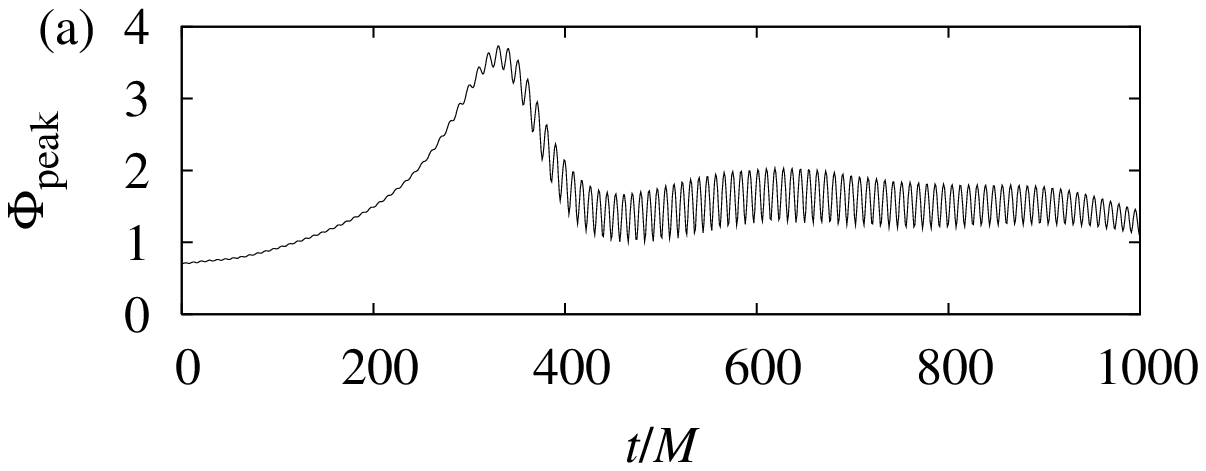}
\includegraphics[width=0.6\textwidth]{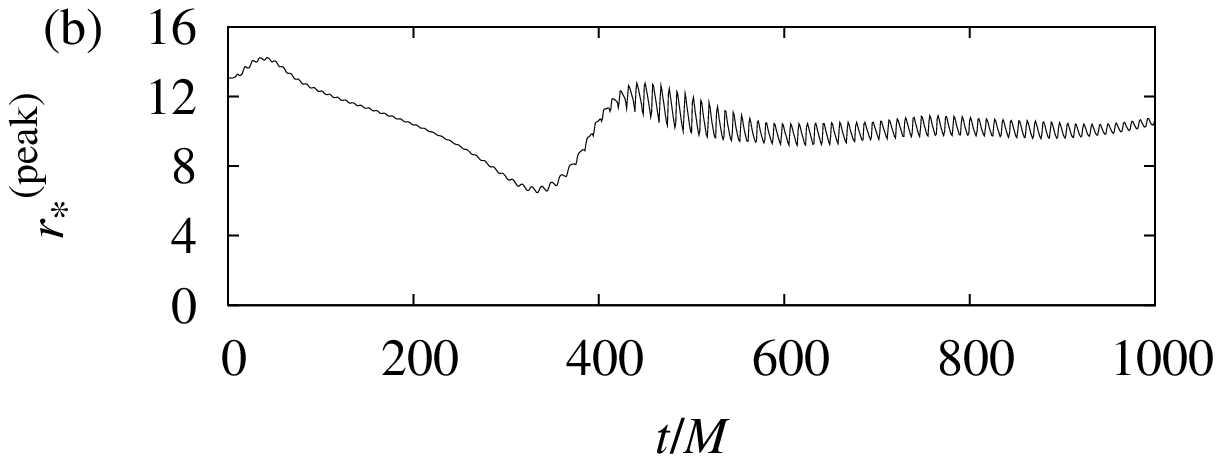}
}
\caption{Same as Fig.~\ref{Fig:amplitude_0.6} but for simulation (B)
[i.e., $\varphi_{\rm peak}(0)=0.7$]. The peak location $r_*^{\rm (peak)}$ 
becomes fairly close to the horizon around 
$t\simeq 350M$, where $\varphi_{\rm peak}$ reaches approximately four.
This is when the bosenova begins to happen, and the behavior after
that time is very different from (A):
$r_*^{\rm (peak)}$ continues small oscillation
around $r_*=10M$ with a short period, and correspondingly, $\varphi_{\rm peak}$
fluctuates around 1.5.
}
\label{Fig:amplitude_0.7}
\end{figure}
%

The upper panel 
of Fig.~\ref{Fig:amplitude_0.7} shows the value of the field
at the peak $\varphi_{\rm peak}=\mathrm{sup}[\varphi]$, 
and the lower panel shows
the position of the peak 
with respect to the tortoise coordinate $r_*$
as functions of $t/M$. In contrast to the case (A),
the value of $\varphi_{\rm peak}$ 
increases only once around $t=300M$, and after that
it fluctuates with short periods.
The position of the peak $r_*^{\rm (peak)}$ also 
approaches the horizon only once, and
after that, it fluctuates around $r_*=10M$ which
is still fairly close to the horizon.

%
\begin{figure}[tb]
\centering
{
\includegraphics[width=0.5\textwidth]{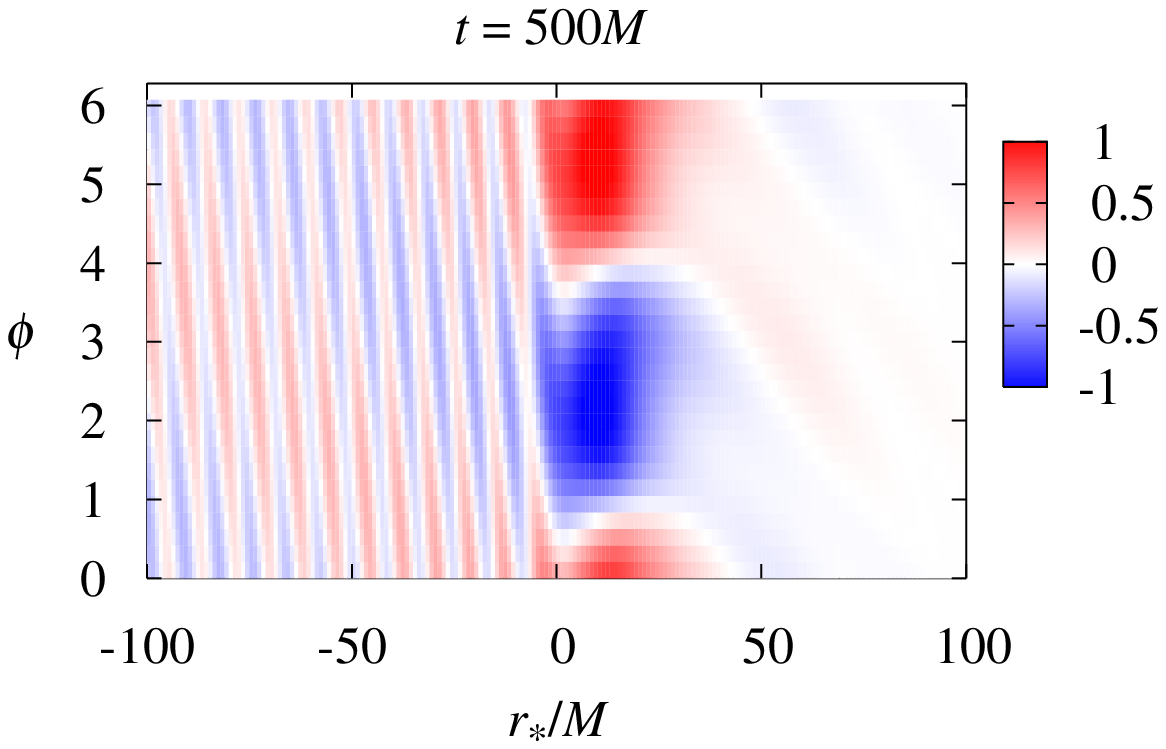}
\includegraphics[width=0.4\textwidth]{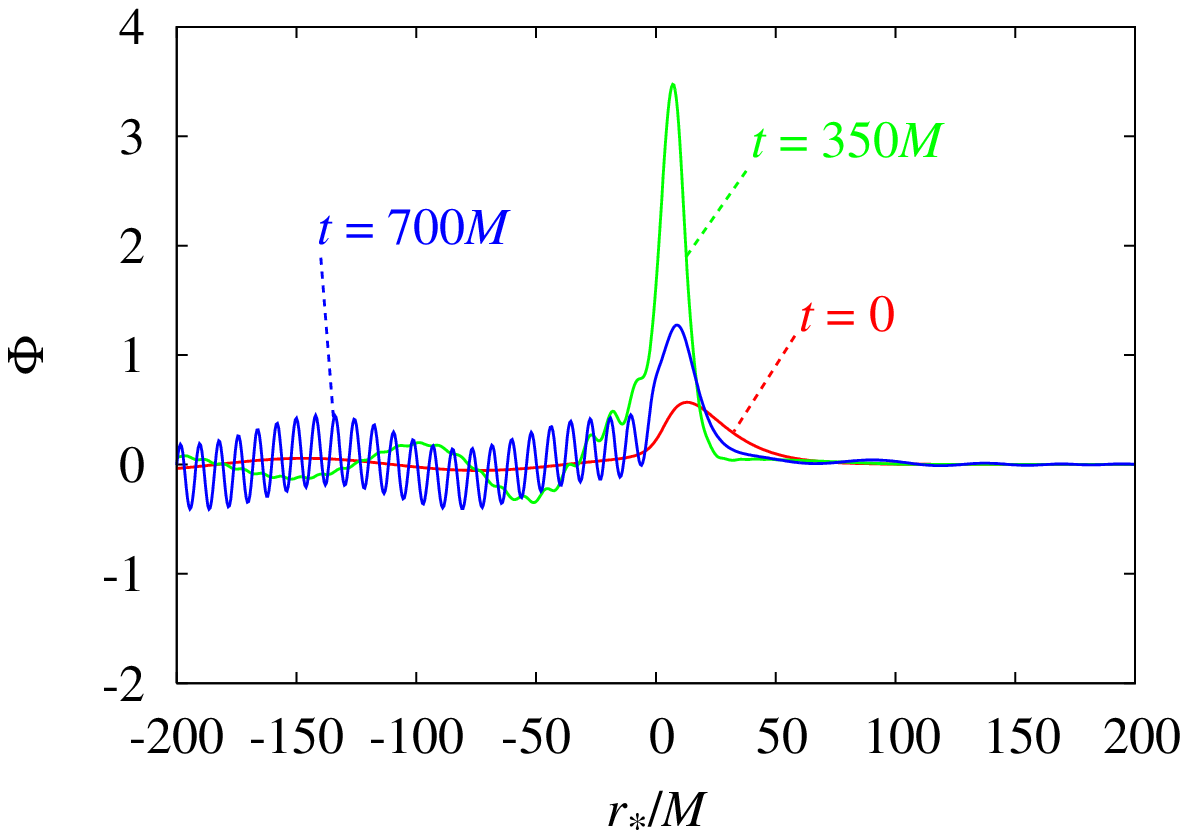}
}
\caption{ 
Left panel: A snapshot of the field 
in the equatorial plane $\theta=\pi/2$
at $t=500M$. Here, the magnitude of the field
$\varphi$ is shown by density plot 
in the plane $(r_*/M, \phi)$.  
Right panel: Snapshots of the field $\varphi$
as a function of $r_*/M$ 
at $\phi=0$ in the equatorial plane $\theta=\pi/2$
for $t/M=0$, $350$, and $700$.
}
\label{Fig:Phi_snapshots}
\end{figure}
%

In order to understand the properties of the bosenova
collapse, let us
look at the field configuration focusing attention to the 
near horizon region. 
The left panel of Fig.~\ref{Fig:Phi_snapshots} shows 
the density plot of the axion field $\varphi$ at $t=500M$
in the $(r_*/M, \phi)$ plane.
The main part of the axion cloud is moving in the $+\phi$ direction.
During the bosenova, ingoing waves that are different from those
of the bound state of the $(\ell,m)=(1,1)$ mode are continuously generated
from the cloud, as can be clearly seen
in this figure. The cloud remains around $r_*\simeq 10M$,
while the waves fall into the BH. 
The generated waves are of the $(\ell, m)=(1,-1)$ mode 
in spite of the fact that the initial condition has just the 
$(\ell, m)=(1,1)$ mode.
The generated waves have 
wavelength $\lambda_{\rm NL}\simeq 8M$ in the tortoise
coordinate $r_*$. This indicates 
$\tilde{\omega}_{\rm NL} \simeq 0.785M$ 
where $\tilde{\omega}:= \omega-m\Omega_H$
(see review in Sec.~\ref{Sec:superradiance}),
and because the waves are in the $m=-1$ mode,
their angular frequency is $\omega_{\rm NL}
\simeq 0.35/M$. 
This angular frequency is approximately same as that of the bound
state of the Klein-Gordon field, $\omega_0\simeq 0.39/M$. 
Since those waves violate the superradiant
condition $\omega<m\Omega_H$ because $m$
is negative, it carries the positive energy toward the horizon.

It is obvious that the generated ingoing waves originate
from the nonlinear effect. However,  
at first glance, the generation of waves of the $(\ell,m)=(1,-1)$ mode
seems strange, because the nonlinear term  
$\sim \varphi_0^3 \sim e^{3i(\phi-\omega_0 t)}$
is unlikely to generate the observed waves
whose behavior is $\sim e^{i(\phi+\omega_{\rm NL} t)}$. 
In Appendix~\ref{Sec:Green}, we study the generation of
waves by the nonlinear effect using the Green's function
approach, and find that waves of the $(\ell,m)=(1,-1)$ mode
actually can be generated. 
Therefore, the waves of the $(\ell,m)=(1,-1)$ mode
found in our simulation are not numerical artifact.
In short, due to the nonlinear effect, several modes of the
bound states (discussed in Sec.~\ref{Sec:superradiance}) of 
frequency $\omega\approx \pm \omega_0$ are 
excited, and these modes include the modes with negative frequency.
Since the $(\ell,m)=(1,1)$ mode of negative frequency is equivalent to the 
$(\ell,m)=(1,-1)$ mode of positive frequency, 
waves of the $(\ell,m)=(1,-1)$ mode can be observed.

The right panel of Fig.~\ref{Fig:Phi_snapshots} shows the
snapshots of the field on the $\phi=0$ line on the equatorial plane
$\theta=\pi/2$, at time $t/M=0$, $350$, and $700$. At 
$t/M=350$, the peak of the field becomes very high, and
around this time, the bosenova begins to happen.
At $t/M=700$, the nonlinear generation of waves of $(\ell,m)=(1,-1)$ mode
continues, 
and the ingoing waves can be seen for $r_*\lesssim 0$.

%
\begin{figure}[tb]
\centering
{
\includegraphics[width=0.9\textwidth]{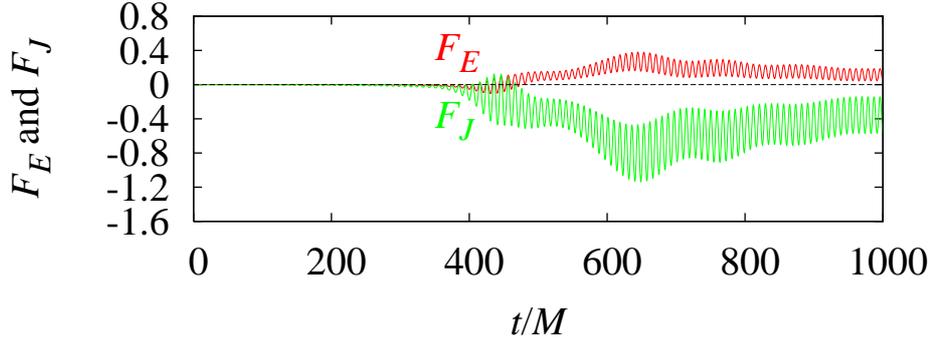}
}
\caption{Same as Fig.~\ref{Fig:EJ-flux-amp0.6}
but for simulation (B)
[i.e., $\varphi_{\rm peak}(0)=0.7$]. 
After the bosenova happens at $t\simeq 350M$, 
the energy flux $F_E$ to the horizon is always positive
while the angular momentum flux $F_J$ is negative. 
Therefore, the energy extraction stops while the
angular momentum extraction continues. 
}
\label{Fig:EJ-flux-amp0.7}
\end{figure}
%

Figure~\ref{Fig:EJ-flux-amp0.7} shows 
the energy flux $F_E$
and the angular momentum flux $F_J$
toward the horizon evaluated at $r_*=-100M$. 
Although both $F_E$ and $F_J$ are negative initially, 
after the bosenova happens,
the value of $F_E$ becomes positive at least up to $t=1000M$. 
Here, the dominant contribution to the flux
comes from the waves of the $(\ell, m)=(1,-1)$
mode generated by the nonlinear effect.
Because those waves  obviously
violate the superradiant condition $\omega<m\Omega_H$,
the energy flux toward the horizon becomes positive. 
As a result, the extraction of energy 
is prevented by the bosenova in this simulation, and 
about 5.3\% of energy falls into the BH 
by $t/M=1500$. 
On the other hand, the value of $F_J$ continues
to be negative, and the waves continue to
extract the angular momentum
from the BH. This is because
the ingoing waves are in the $m=-1$ mode,
and hence, carry negative angular momentum if energy is positive. 
The small fluctuations of $F_E$ and $F_J$ come from the interference
of the $(\ell, m)=(1,\pm 1)$ modes, and 
the typical angular frequency of this oscillation is $\approx 2\omega_0$,
similarly to the simulation (A).

%
\begin{figure}[tb]
\centering
{
\includegraphics[width=0.45\textwidth]{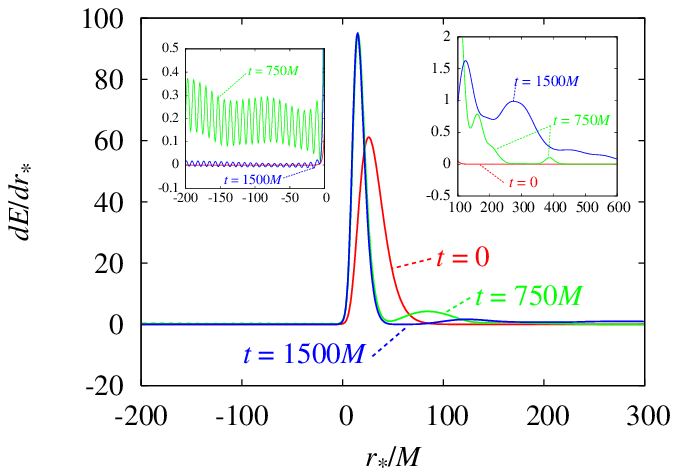}
\includegraphics[width=0.45\textwidth]{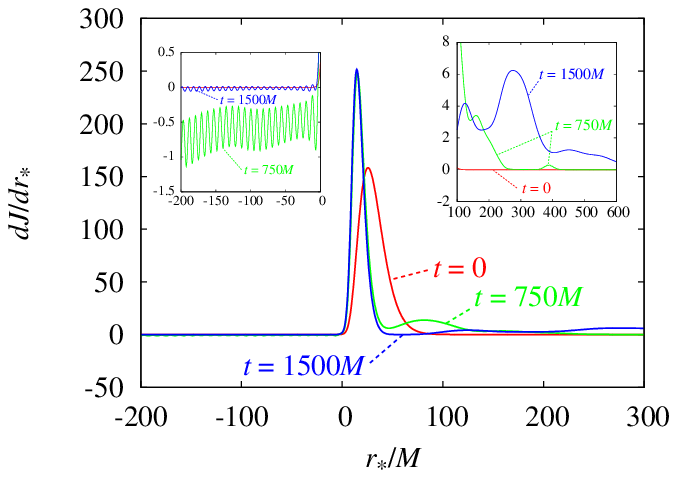}
}
\caption{The energy density $dE/dr_*$ (left) and the angular momentum 
density $dJ/dr_*$ (right) with respect to the tortoise coordinate $r_*$
at time $t/M=0$, $750$, and $1500$ for simulation (B)
[i.e., $\varphi_{\rm peak}(0)=0.7$]. 
}
\label{Fig:dEdr-dJdr-amp0.7}
\end{figure}
%

The left and right panels 
of Fig.~\ref{Fig:dEdr-dJdr-amp0.7} show the energy density
$dE/dr_*$ and the angular momentum density $dJ/dr_*$
with respect to the tortoise coordinate $r_*$, respectively, 
at $t/M=0$, $750$, and $1500$. At late time,
most of the energy is contained 
in the domain $0\lesssim r_*/M\lesssim 30$, and the
energy and angular momentum densities have similar shapes
for $t/M=750$ and $1500$. The difference between
$t/M=750$ and $1500$ can be seen in the
domains $r_*/M\lesssim 0$ and $r_*/M\gtrsim 100$. 
The behavior of each of $dE/dr_*$ and $dJ/dr_*$
in the domain $r_*/M\lesssim 0$
can be seen in the left inset of each panel.
At $t=750M$, the energy density is positive and 
the angular momentum density is negative. This is consistent
with the fact that the energy and angular momentum fluxes
to the horizon are positive and negative, respectively, as seen in
Fig.~\ref{Fig:EJ-flux-amp0.7}.
At $t=1500M$, both two densities fluctuate around zero,
and the mean values of energy and
angular momentum densities are still positive and negative,
respectively. This is because the nonlinear resonance
becomes weak at this time, and the system settles down 
to a quasistationary state again. 
The behavior of each of $dE/dr_*$ and $dJ/dr_*$ at the distant place
is shown in the right inset of each panel. 
At $t=750M$, some fraction of energy and angular momentum are
distributed at a far region. Around $r_*=400M$, 
a small bump can be seen. This bump moves outward
approximately at the speed of light. Therefore, a kind of
``explosion'' happens in the bosenova. 
However, this explosion is very small because this bump has
only $\approx 0.2$\% of the total energy. 
At $t=1500M$, more amount of
energy and angular momentum can be seen at the distant place.
Therefore, following the small explosion, the field energy gradually
spreads out to the distant region. 
At $t/M=1500$, about $16.6\%$ of the total energy
distributes in the region $r_*/M\ge 60$. Except for the small bump
moving at the speed of light, all field in the distant region 
seems to be gravitationally
bounded. The simulation was performed up to $t/M=2000$, and
it was found that after $t/M=1500$, some part of the energy 
at the distant place begins to fall back.

%
\begin{figure}[tb]
\centering
{
\includegraphics[width=0.6\textwidth]{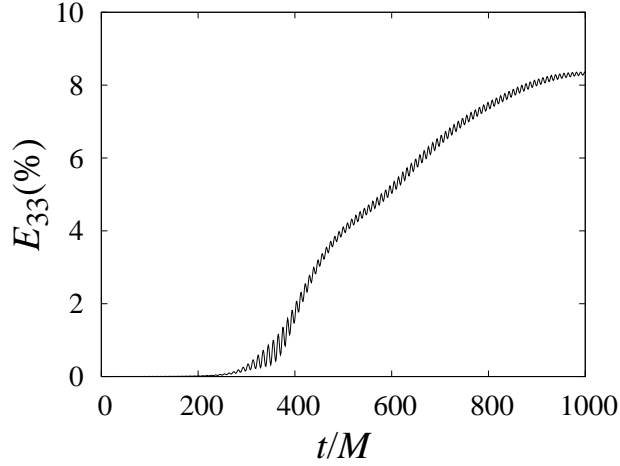}
}
\caption{The estimated amount of energy $E_{33}$ of $(\ell,m)=(3,\pm 3)$ mode 
generated by the mode mixing in the bosenova 
as a function of time $t/M$. Here $E_{33}$ 
is normalized by the total energy $E$ and shown in the unit of \%.
About 10\% energy is converted into the $(\ell,m)=(3,\pm 3)$ mode.}
\label{Fig:energy33}
\end{figure}
%

It is interesting to study how much energy of the
axion field is converted from the $(\ell,m)=(1,1)$ mode
to other modes. Unfortunately, 
numerical decomposition of the field $\varphi$ into the
modes is rather tedious because it requires
Fourier transform from the time domain to the frequency domain, and also, 
in the case of the Kerr spacetime, the separation 
constant and the spheroidal harmonics
depend on the frequency $\omega$.
Instead,  
we give rough estimate on how much energy
is converted to the $(\ell,m)=(3,\pm 3)$ mode 
in an approximate way. In this approximation,  
we use the spherical harmonics instead of 
the spheroidal harmonics, and decompose the field into the
form $\varphi=\sum_{0\le m\le\ell}\varphi_{\ell m}$,
where $\varphi_{\ell m}:=f_{\ell m}(t,r)Y_{\ell m}(\theta,\phi)
+f_{\ell -m}(t,r)Y_{\ell -m}(\theta,\phi)$.
The energy $E_{\ell m}$ of each $(\ell,\pm m)$ mode is estimated
by substituting $\varphi_{\ell m}$ into the formula for the energy.
In this manner, we have studied the mode $0\le l\le 4$,
and found that the modes with $(\ell, m)=(1,\pm 1), (3,\pm 1)$,
and $(3,\pm 3)$ 
are nonzero and the other modes are approximately zero.

Figure~\ref{Fig:energy33} shows the value of energy of
the $(\ell, m)=(3, \pm 3)$ mode normalized by the total energy,
$E_{33}/E$, as a function of time. After the bosenova
happens, the $(\ell, m)= (3, \pm 3)$ mode starts to grow 
and has about 10\% of the total
energy at $t/M=1000$. 
On the other hand, before the bosenova,
the mode mixing is fairly weak, and almost all fields
are in the $(\ell, m)=(1,1)$ mode. 
We evaluated the value of $E_{33}/E$
also for the simulation (A) [i.e., $\varphi_{\rm peak}(0)=0.6$], 
and found that the $(\ell, m)=(3,\pm 3)$ mode has at most $0.014$\% 
of the total energy.
This result shows that the bosenova converts relatively large amount
of the axion field from the $(\ell, m)=(1,1)$ mode to other modes.

To summarize, if the initial amplitude is sufficiently large,
the bosenova happens for the axion cloud of quasibound state
of the $(\ell,m)=(1,1)$ mode. The bosenova
in this system is characterized by the following features.
First, a small amount ($\approx 0.2$\%) of energy comes out 
from the axion cloud approximately at the speed of light. 
After that, about $15$\% of energy gradually spreads out
to the distant region, although it seems to be gravitationally
bounded. Therefore, the bosenova of the BH-axion
system is somewhat similar to the bosenova of BEC in experiments
(In Sec.~\ref{Sec:summary}, we give a more detailed comparison). 
Second, about $5$\% of energy
falls into the BH after the bosenova. This energy
is carried by waves of the $(\ell, m)=(1,-1)$ mode generated
by the nonlinear effect. Therefore, in the BH-axion system,
the ``explosion'' happens both to distant place and to the horizon. 
Finally, once the bosenova happens, the mode mixing effectively
occur. In addition to the generation of the $(\ell,m)=(1,-1)$ mode,
the $\ell=3$ mode was observed to get $\approx 10\%$ 
of the total energy after the bosenova.

\subsection{Does the bosenova really happen?}
\label{Sec:bosenova}

In the two simulations performed above, we 
found the following:
In the simulation (A), when the initial peak value is small, 
the nonlinear effect causes 
periodic changes in the peak location and the peak value,  
and enhances the energy and angular momentum 
extractions; In the simulation (B), when the initial peak value
is large, the nonlinear effect
causes an explosion, the bosenova. 
During the bosenova, waves of the $(\ell,m)=(1,-1)$ mode violating
the superradiant condition are generated, and thus, the energy flux 
toward the horizon becomes positive
terminating the superradiant instability.

%
\begin{figure}[tb]
\centering
{
\includegraphics[width=0.5\textwidth]{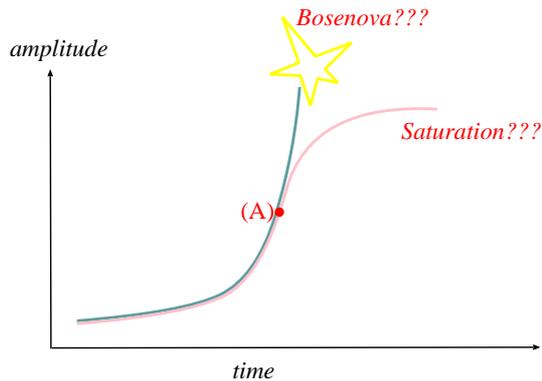}
}
\caption{
Schematic picture of time evolution of the field amplitude
and two possibilities of the final state of the superradiant instability.
}
\label{Fig:evolution}
\end{figure}
%

In this subsection,
we discuss whether the bosenova happens as the result
of the superradiant instability. 
In a realistic system, as the rotational energy of the
BH is extracted, the field gradually 
gets amplified. In this sense, 
the simulation (B) may be artificial because we gave 
a quasibound state of large amplitude by hand and
its initial condition may not be realized as the
result of the superradiant instability.  
In particular, we have to take care of the possibility that
the bosenova does not happen,
as there are a lot of examples of nonlinear systems in which
the nonlinear effects saturate the instabilities leading
the systems to quasistable states. 
Figure~\ref{Fig:evolution} is a schematic picture
depicting the two possibilities of time evolution: 
the bosenova and the saturation.

In order to discuss which is the case,
we perform supplementary simulations as follows. 
In these simulations, we prepare the initial condition
by applying the scale transformation
to the result of the simulation (A) at $t=1000M$ 
as $\varphi(0) = C\varphi^{(A)}(1000M)$ 
and $\dot{\varphi}(0) = C\dot{\varphi}^{(A)}(1000M)$.
This is because in the presence of the nonlinear term,
the energy gets confined in a
smaller region near the BH compared to the case of the
quasibound state of the massive Klein-Gordon field
as found in Fig.~\ref{Fig:dEdr-dJdr-amp0.6}, and 
the initial condition prepared by this procedure
is expected to be more realistic and to approximate
the nearly final state of the superradiant instability. 
The cases of $C=1.05$, $1.08$, and $1.09$
were simulated. Here, 
the initial state for a larger $C$ is expected to approximate
the later state of Fig.~\ref{Fig:evolution}.

%
\begin{figure}[tb]
\centering
{
\includegraphics[width=0.7\textwidth]{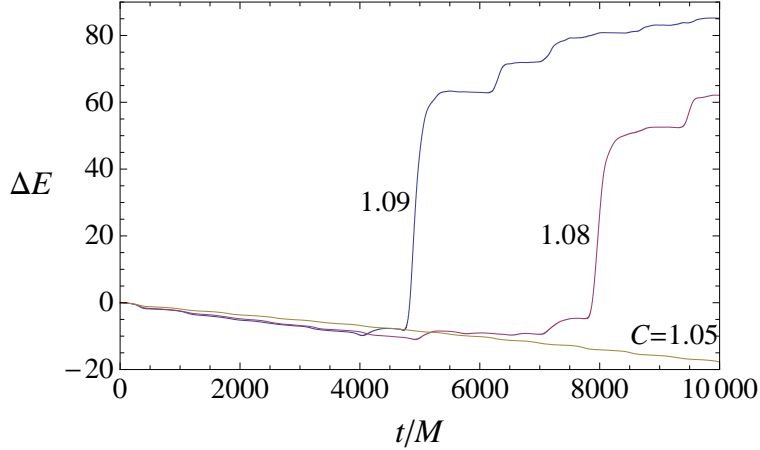}
}
\caption{The relation between time $t$ and
the amount of energy $\Delta E$ that has fallen into the
BH by the time $t$. Here, the cases of $C=1.05$, $1.08$, 
and $1.09$ are shown.
}
\label{Fig:integrated-E}
\end{figure}
%

Figure~\ref{Fig:integrated-E} shows the relation between
time $t$ and the amount energy $\Delta E$ that
has fallen by the time $t$. In other words, 
$\Delta E$ is the integrated energy flux toward the
horizon from time zero to $t$:  $\Delta E:=\int_0^t F_Edt$. 
The gradient of each curve shows the flux $F_E$ toward the horizon:
If it is negative (resp. positive), 
the negative (resp. positive) energy is falling into the BH.
The curve of the case $C=1.05$ is always negative, 
and at this stage, the energy continues to be 
extracted. In this case, the averaged rate of the superradiant 
instability is $\gamma_{\rm NL} M \approx 5.85\times 10^{-7}$, which
is larger than that of the case of 
the linear Klein-Gordon field, $\gamma M=1.30\times 10^{-7}$.
This confirms
that the nonlinear effect enhances the 
rate of superradiant instability before the bosenova.
The bosenova does not happen at least
by time $t/M=10000$. 

Next, let us look at the case $C=1.09$. In this case,
the value of $F_E$ is negative until $t/M\simeq 4000$.
Here, the rate of energy extraction is 
$\gamma_{\rm NL} M\simeq 7.45\times 10^{-7}$, which is further
larger than that of the case $C=1.05$.
However, around $t/M\simeq 4000$, the value of $F_E$ 
becomes positive, and for $t/M\gtrsim 4500$, the value of 
$F_E$ becomes positive and fairly large: the
bosenova happens around this time.
This case represents the example such that the bosenova
happens after a certain period of energy extraction. Therefore,
it is natural to consider that this simulation approximates 
what actually happens at the final stage of the
superradiant instability. 
The burst of positive energy toward the horizon continues
until $t/M\simeq 5300$, and after that small explosions
happen intermittently. 

In the case of $C=1.08$, the energy extraction continues
from $t/M=0$ to $5000$. Around $t/M=5000$ and $7500$, 
small amounts of positive energy fall into the BH, and then,
around $t/M=8000$, a large positive ingoing
energy flux is generated. This is the bosenova in this case.
The bosenova in the case $C=1.08$ happens later than the case 
$C=1.09$, mainly because the initial amount of energy of the former
is smaller than that of the latter, and therefore, a longer
period of energy extraction is required.
However, it should be noted
that the bosenova of these two cases happen at different 
values of energy: The amount of energy 
when the positive flux is first generated is 
 $E/[(f_a/M_p)^2M]\simeq 1633$ and $1607$ 
for the cases $C=1.09$ and
$1.08$, respectively. Although the criterion for the occurrence
of the bosenova is mainly determined by the energy amount,
it would depend also on the detailed structure
of the axion cloud. 

The natural picture of the final stage of the superradiant
instability is as follows. Before the bosenova, 
similarly to the case of $C=1.05$, the energy
continues to be extracted and the rate of the energy extraction
is gradually enhanced. Then, at a certain critical 
point, where the energy of the axion cloud is $E/[(f_a/M_p)^2M]\approx 1620$,
a large amount of positive energy suddenly falls into the BH,
and here, the bosenova happens like the simulations of $C=1.08$ and $1.09$. 
The answer to the question ``Does the bosenova really happen?'' 
is ``{\it Yes},'' because
from our simulations,
it is natural to consider that the bosenova actually happens
as a result of the superradiant instability. 
In particular,
we have obtained no evidence for the possibility
that the nonlinear effect saturates the growth by the superradiant
instability and leads 
the system to a quasistable state.

If we assume the decay constant $f_a$ to be the GUT scale,
$\approx 10^{16}$GeV, the bosenova collapse happens when the
energy of the axion cloud grows to be $E\simeq 1.6\times 10^{-3}M$,
i.e., when the axion cloud gets energy of $\approx 0.16$\% of the
BH mass. Therefore, if the value of $f_a$ is the GUT scale
or smaller, the back reaction to the background spacetime, 
such as the change in the parameter
in $M$ and $a$ of the BH or the distortion of the background geometry
by the axion cloud, is negligible. 
On the other hand, if we assume $f_a\approx 10^{17}$GeV, the
bosenova collapse happens when $E\simeq 0.16M$, i.e., 
when the axion cloud gets energy of $\approx 16$\% of the BH mass.
In such a situation, back reaction to the background geometry
is significant, and the bosenova phenomena has to be studied
by the method of numerical relativity.

The time evolution long after the bosenova is also an interesting
subject, although this is beyond the scope of this paper. Looking
at Fig.~\ref{Fig:integrated-E}, in the case of $C=1.09$, the small amount
of positive energy intermittently falls into the BH. 
One possibility is that this phenomena continues
and the axion cloud continue to extract
and lose small amounts of energy.
Another possibility is that the axion cloud loses almost all energy,
and the superradiant instability happens from the beginning,
leading to the bosenova collapse that has approximately  
same scale as the previous one. In order to clarify which is the case,
a very-long-term simulation or construction of a good approximate
model is necessary.

When we consider a very-long-term evolution
of the BH-axion system, taking account of changes in mass 
$M$ and angular momentum $J$ of the BH is also important.
In our simulations, the energy is extracted from the BH in superradiant
instability and falls back to the BH in the bosenova collapse.
On the other hand, the angular momentum is extracted both in the
superradiant instability and in the bosenova collapse. Therefore,
the spin parameter $a/M$ would gradually decrease
in a very-long-term evolution. As a result, superradiant 
instability of the $(\ell,m)=(1,1)$ mode will stop when $a/M$ is decreased
to a certain value that is determined by the mass $\mu$ of the axion
(see Fig.~7 of Ref.~\citen{Dolan:2007}).
After that, superradiant instability
of the $(\ell,m)=(2,2)$ mode will become a primary factor in
determining the evolution of the axion cloud.

%
%
\section{Effective theory of axion cloud model}
\label{Sec:effective_theory}

In this section, we discuss the reason why the bosenova
happens in the BH-axion system by introducing an effective  
theory for this system. In this discussion, we model the
axion cloud using a time-dependent Gaussian
wavefunction. Also, we assume the
non-relativistic approximation in which  
the gravity is treated by the Newtonian potential. The distribution
is specified by the following three parameters: $\delta_r$
(the width of the wavepacket
along the $r$ direction), 
$\delta_\nu$  (the width
along the $\theta$ direction), and
$r_p$ (the position of the peak with respect to the $r$ coordinate). 
Under these approximations, we derive the  
effective action for the three parameters and find
various properties of the bosenova which are consistent
with the simulations.

\subsection{Effective action}

The action for the axion field $\Phi$ is given by Eq.~\eqref{Eq:action}.
In terms of the normalized field $\varphi=\Phi/f_a$, the action is
rewritten as
\begin{equation}
\hat{S} = \int d^4x \sqrt{-g}
\left[-\frac12 (\nabla\varphi)^2 -\mu^2
\left(\frac{\varphi^2}{2}+
\hat{U}_{\rm NL}(\varphi)\right)\right],
\label{Eq:action2}
\end{equation}
where the nonlinear potential $\hat{U}_{\rm NL}$ is defined by
\begin{equation}
\hat{U}_{\rm NL}(\varphi) = 1-\frac{\varphi^2}{2}-\cos\varphi
=-\sum_{n=2}^{\infty}\frac{(-1)^n}{(2n)!}x^{2n}.
\end{equation}
We introduce $\psi$ as
\begin{equation}
\varphi = \frac{1}{\sqrt{2\mu}}
\left(e^{-i\mu t}\psi + e^{i\mu t}\psi^*\right).
\end{equation}
Here, $\psi$ is a slowly varying function under the non-relativistic
approximation. Substituting this formula into Eq.~\eqref{Eq:action2}, 
we have
\begin{equation}
\hat{S}_{\rm NR} = \int d^4x
\left[\frac{i}{2}\left(\psi^*\dot{\psi}-\psi\dot{\psi}^*\right)
-\frac{1}{2\mu}\partial_i\psi\partial_i\psi^*
+\frac{\alpha_g}{r}\psi^*\psi
-\mu^2\tilde{U}_{\rm NL}(|\psi|^2/\mu)
\right],
\label{Eq:action_NR}
\end{equation}
where $\alpha_g:=M\mu$ and 
\begin{equation}
\tilde{U}_{\rm NL}(x) = -\sum_{n=2}^{\infty}
\frac{(-1/2)^n}{(n!)^2}x^n.
\end{equation}
Here, the Newtonian approximation is adopted for gravity.

Now, we assume the form of $\psi$ as 
\begin{equation}
\psi = A(t,r,\nu)e^{iS(t,r,\nu)+m\phi},
\end{equation}
where $\nu:=\cos\theta$ and we set $m=1$. The functions $A(t,r,\nu)$
and $S(t,r,\nu)$ are chosen to be the following form: 
\begin{equation}
A(t,r,\nu) \approx A_0 \exp\left[-\frac{(r-r_p)^2}{4\delta_r r_p^2}
-\frac{(\nu-\nu_p)^2}{4\delta_\nu}\right],
\label{Eq:A_formula}
\end{equation}
\begin{equation}
S(t,r,\nu) \approx
S_0(t) + p(t)(r-r_p) + P(t)(r-r_p)^2 + \pi_\nu(t) (\nu-\nu_p)^2+\cdots.
\end{equation}
$\delta_r(t)$ is the width of the wavepacket along the $r$
direction, $\delta_\nu(t)$ is the width along the $\nu$ direction 
(i.e., $\theta$ direction), 
and $r_p(t)$ is the position of the peak with respect to $r$ coordinate. 
Since the center of the wavepacket
always exists on the equatorial plane,
the peak position with respect to $\nu$ coordinate
is always zero, $\nu_p\equiv 0$. We define $N$ as
\begin{equation}
N=\int d^3x A^2 \approx 4\pi^2A_0^2\sqrt{\delta_r\delta_\nu}r_p^3(1+\delta_r).
\end{equation}
Here, we ignored the inner cutoff of the integration range 
of $r$. In a similar manner, we perform the integration 
of the action (with respect to spatial coordinates) and 
derive the Lagrangian density as
\begin{equation}
L=-\dot{S}_0N + p\dot{r}_pN 
+(\dot{p}-2P\dot{r}_p)2r_p\frac{\delta_r}{1+\delta_r}N
-\dot{P}r_p^2\delta_r\frac{1+3\delta_r}{1+\delta_r}N - \dot\pi_\nu \delta_\nu N
-H
\end{equation}
with 
\begin{equation}
H=T+V,
\end{equation}
where
\begin{equation}
T = \frac{N}{2\mu}
\left[
p^2 
+8pPr_p\frac{\delta_r}{1+\delta_r}
+ 4P^2r_p^2\delta_r\frac{1+3\delta_r}{1+\delta_r}
+4\pi_\nu^2\frac{\delta_\nu}{r_p^2(1+\delta_r)}
\right]
\end{equation}
\begin{multline}
\frac{V}{N\mu\alpha_g^2}
=\frac{1}{2(\alpha_g\mu r_p)^2(1+\delta_r)}
\left(1+\delta_\nu
+\frac{1}{4\delta_r}+\frac{1}{4\delta_\nu}
\right)
-\frac{1}{(\alpha_g\mu r_p)(1+\delta_r)}
\\
-\alpha_g^{-2}\sum_{n=2}^{\infty}
\frac{(-1/2)^n}{(n!)^2n}
\left[\frac{N_*}{\sqrt{\delta_r\delta_\nu}
(\alpha_g\mu r_p)^3(1+\delta_r)}\right]^{n-1},
\label{Eq:potential}
\end{multline}
with $N_* = (\alpha_g^3\mu^2/4\pi^2)N$. 
Here, we have kept terms up to the first-order
in $\delta_\nu$ because the wavepacket form, Eq~\eqref{Eq:A_formula},
is not a very good approximation in the $\nu$ direction and 
taking account of higher-order terms in $\delta_\nu$ 
does not have an important meaning. 
The term depending on $\dot{S}_0$ can be omitted 
since it just gives the conservation of $N$. The variables
$p$, $P$, and $\pi_\nu$ can be related to the 
conjugate momenta of the variables $\delta_r$, 
$\delta_\nu$, and $r_p$, and therefore, we can express
the Lagrangian only in terms of $\delta_r$, $\delta_\nu$, 
and $r_p$, and their time derivatives. To summarize, 
the equivalent Lagrangian is
\begin{equation}
L=T-V,
\end{equation}
\begin{equation}
T = \frac12 A\dot{\delta}_r^2 
+ B\dot{\delta}_r \dot{r}_p
+ \frac12 C \dot{r}_p^2
+ \frac{1}{2}D
\dot{\delta}_\nu^2,
\end{equation}
where
\begin{subequations}
\begin{equation}
A=\frac14 N\mu r_p^2
\frac{1+45\delta_r+198\delta_r^2+126\delta_r^3
+45\delta_r^4 + 9\delta_r^5}
{(1+\delta_r)^3\delta_r(1+3\delta_r^2)},
\end{equation}
\begin{equation}
B=\frac12 N\mu r_p\frac{-7-30\delta_r+54\delta_r^2+30\delta_r^3
+9\delta_r^4}{(1+\delta_r)^2(1+3\delta_r^2)},
\end{equation}
\begin{equation}
C=N\mu\frac{1+6\delta_r-26\delta_r^2+18\delta_r^3
+9\delta_r^4}{(1+\delta_r)(1+3\delta_r^2)},
\end{equation}
\begin{equation}
D = \frac{1}{4}N\mu r_p^2\frac{(1+\delta_r)}{\delta_\nu}.
\end{equation}
\end{subequations}
The potential $V$ is given in Eq.~\eqref{Eq:potential}.

\subsection{Equilibrium point of the potential}

\label{Sec:equilibrium_point}

Let us study the properties of the potential $V$. 
This potential is a function defined in 
the three-dimensional phase space
$(\delta_r, \delta_\nu, \alpha_g\mu r_p)$. 
For each $N_*$, 
the equilibrium point is determined by
$V_{,\delta_\nu}=V_{,\delta_r}=V_{,r_p}=0$.
From this equilibrium condition, the 
following two simple relations are obtained:
\begin{equation}
\delta_r = \frac
{-1+4\delta_\nu^2+
\sqrt{1-8\delta_\nu+8\delta_\nu^2+64\delta_\nu^3+16\delta_\nu^4}}
{2(-2+4\delta_\nu+16\delta_\nu^2)},
\label{Eq:deltar_deltanu}
\end{equation}
\begin{equation}
\alpha_g\mu r_p
=4\delta_\nu -\frac{1}{2\delta_\nu}
+\frac{1}{4\delta_r}+1.
\label{Eq:rp_deltanu}
\end{equation}
If we impose these two conditions \eqref{Eq:deltar_deltanu}
and \eqref{Eq:rp_deltanu}, 
we have a line in the three-dimensional phase space 
$(\delta_r, \delta_\nu, \alpha_g\mu r_p)$, and along this line,
the potential $V$ can be regarded as the function of
$\alpha_g\mu r_p$. Let us study the behavior of this function 
$V(\alpha_g\mu r_p)$
varying the value of $N_*$.
We also plot the value of $\alpha_g\mu r_p$ at the 
equilibrium point as a function of $N_*$. In the following, 
the two cases $\alpha_g=0.1$ and $0.4$ are shown.

\subsubsection{The case $\alpha_g=0.1$}
\label{Sec:equilibrium_0.1}

%
\begin{figure}[tb]
\centering
{
\includegraphics[width=0.45\textwidth]{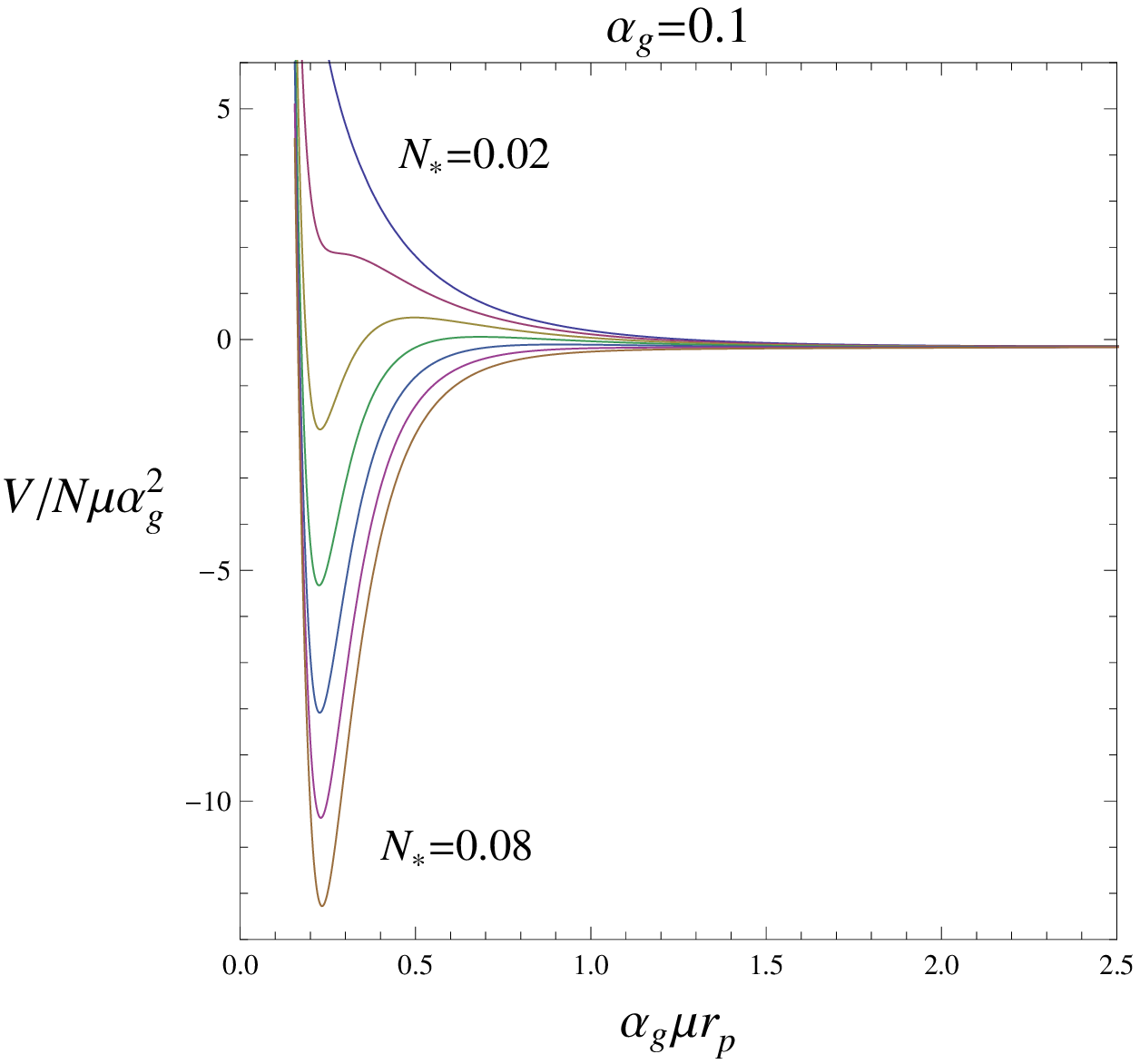}
\hspace{5mm}
\includegraphics[width=0.45\textwidth]{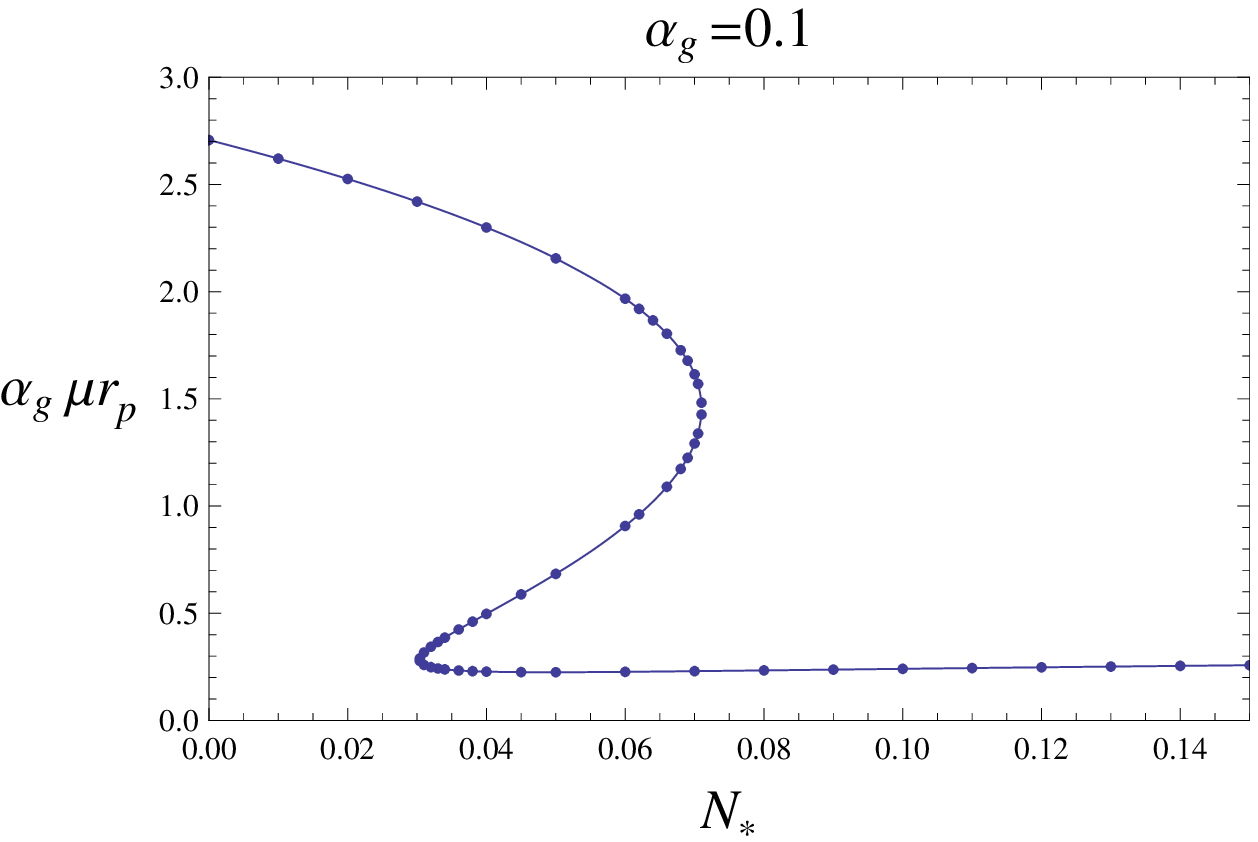}
}
\caption{Left panel: The behavior of the potential $V$ as a function
of $\alpha_g\mu r_p$ for $N_*=0.02$,...,$0.08$ with $0.01$ intervals
in the case $\alpha_g=0.1$.
There is one stable point for $N_*=0.02$, and as $N_*$ is increased, 
another stable point appears for $N_*\approx 0.03$. 
The outer stable point disappears for $N_*\approx 0.07$ and the
system moves to the inner stable point (i.e., the bosenova). 
For $N_*= 0.08$, there is only one stable point at 
$\alpha_g\mu r_p\approx 0.025$. 
Right panel: The position of the equilibrium point as a function
of $N_*$ in the case $\alpha_g=0.1$. For $0.0304\lesssim N_*\lesssim 0.071$, 
there are two stable points and one unstable point, while
only one stable point exists for $N_*\lesssim 0.0304$ and
$\gtrsim 0.071$.}
\label{Fig:peaklocation_alpha01}
\end{figure}
%

Let us look at the case $\alpha_g=0.1$. 
The left panel of Fig.~\ref{Fig:peaklocation_alpha01} 
shows the relation between $\alpha_g\mu r_p$ and 
$V$ along the line introduced above. The cases of 
$N_* =0.02$,...,$0.08$ are depicted with $0.01$ intervals.  
The right panel of Fig.~\ref{Fig:peaklocation_alpha01} 
shows the relation between $N_*$
and $\alpha_g\mu r_p$ at the equilibrium point.   For 
$N_*<0.0304$, there is only one stable point.
But as the $N_*$ is increased, the situation changes: 
At $N_*\simeq 0.0304$, another stable point appears inside of the 
original stable point,
and the two stable points exist for $0.0304\lesssim N_*\lesssim 0.071$.
At $N_*\simeq 0.071$, the outer stable point disappear, and 
only one stable point exists for $N_*\gtrsim 0.071$.

The interpretation of this result is as follows. Because of the
superradiant instability, the value of $N_*$ is gradually increased
and the shape of the potential gradually changes. The 
position of the wavepacket is around $\alpha_g\mu r_p\approx 2.7$
for small $N_*$, and it becomes smaller as $N_*$ is increased.
When $N_*$ reaches $\approx 0.03$, a new stable point appears inside 
of the original stable point.  The system remains at the original
outer stable point for a while, i.e., until $N_*$ reaches $\approx 0.07$.
When $N_*$ exceeds $\approx 0.07$, the outer stable point disappears,
and therefore, the system jumps from the original outer stable point
to the inner stable point. Accordingly, the value of $\alpha_g\mu r_p$ 
jumps from $\approx 1.5$ to $\approx 0.25$. So, the phase transition
occurs when the amplitude reaches some critical point, 
and this is interpreted as the bosenova collapse. This is consistent
with our simulation results: In simulation (A), the peak position
continues to oscillate in relatively distant region, while in
simulation (B), the peak position remains in the neighborhood
of the horizon after the bosenova.

\subsubsection{The case $\alpha_g=0.4$}

\label{Sec:equilibrium_0.4}

%
\begin{figure}[tb]
\centering
{
\includegraphics[width=0.45\textwidth]{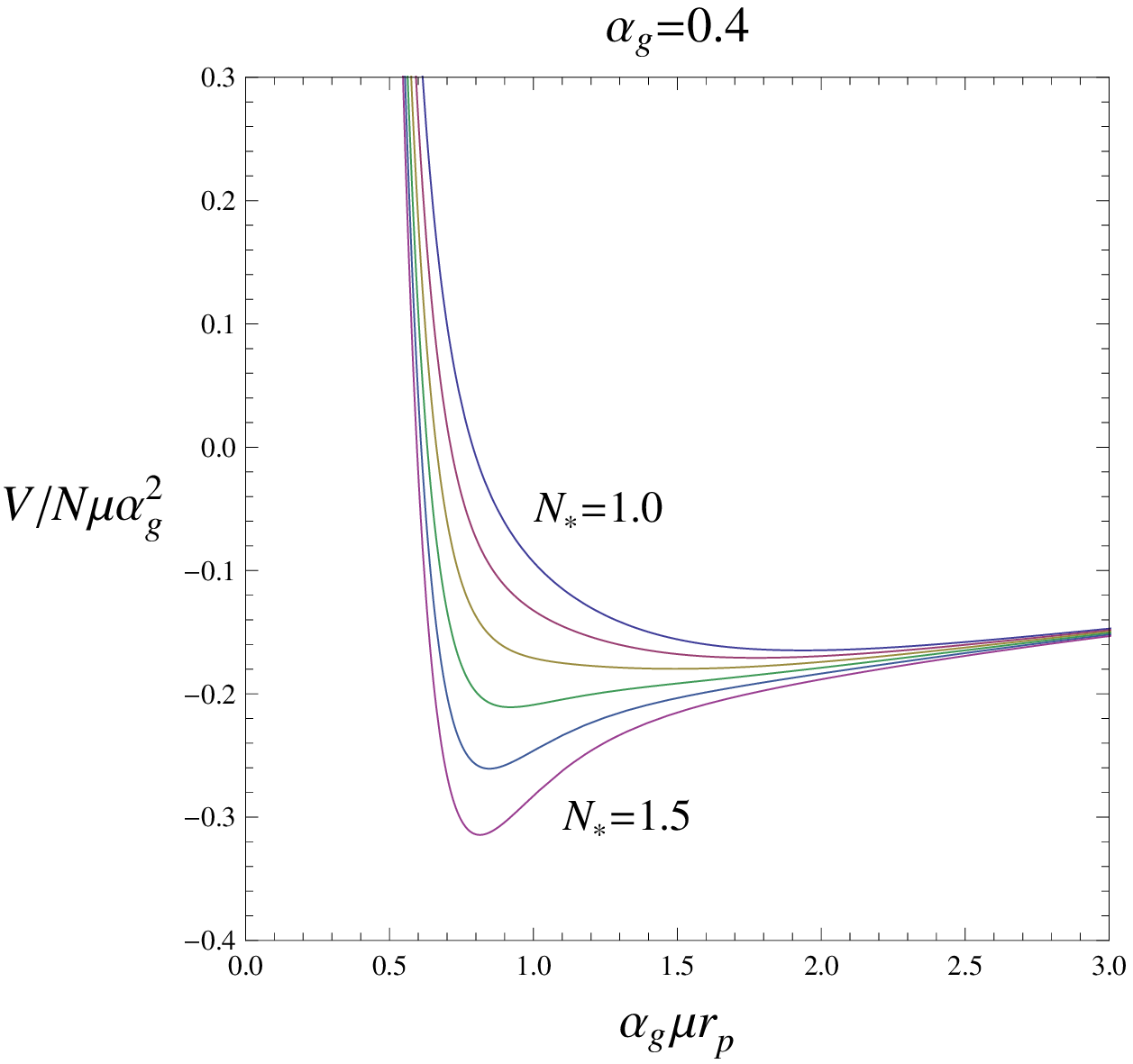}
\hspace{5mm}
\includegraphics[width=0.45\textwidth]{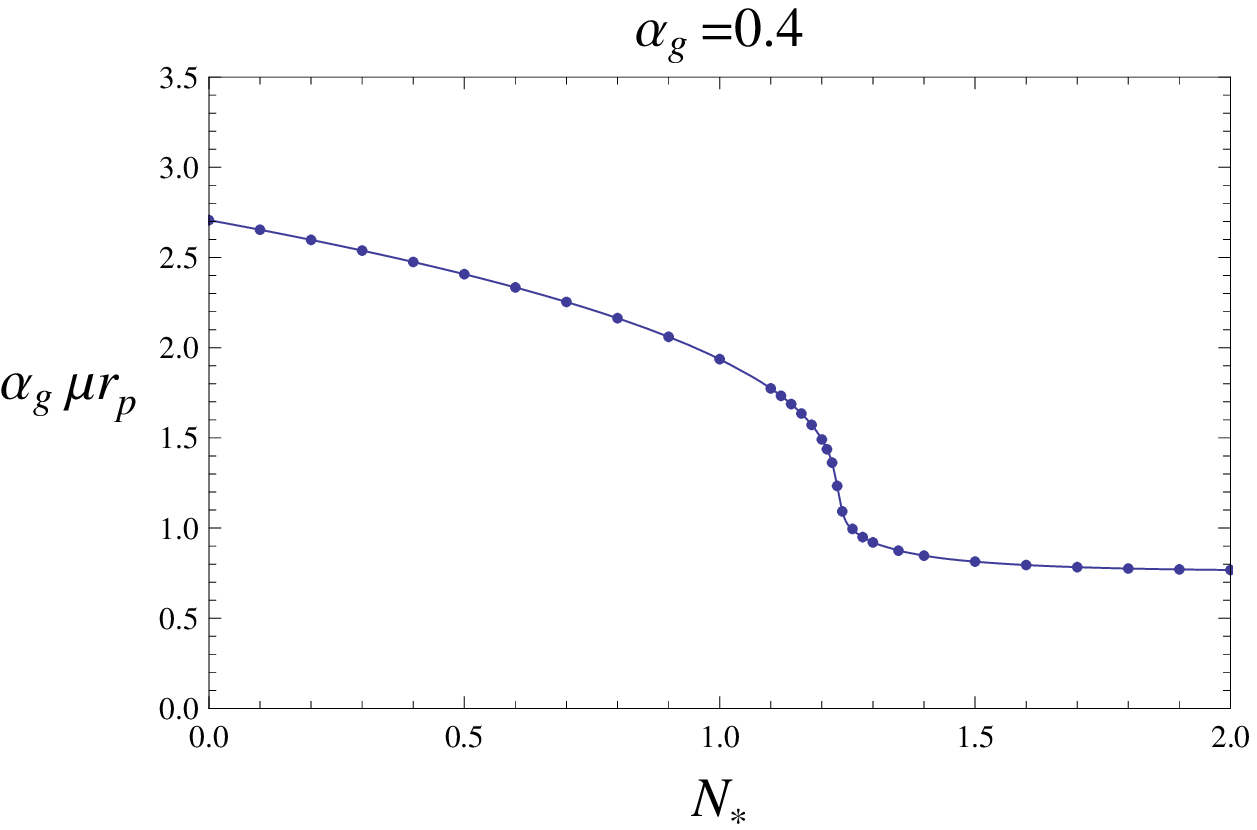}
}
\caption{The same as Fig.~\ref{Fig:peaklocation_alpha01} but
for $\alpha_g=0.4$. The cases $N_*=1.0$,...,$1.5$ are shown
with $0.1$ intervals for left panel.
In this case, only one stable point exists
for all $N_*$. Around $N_*=1.2$, the position of the peak 
$\alpha_g\mu r_p$ rapidly changes
to a smaller value. }
\label{Fig:peaklocation_alpha04}
\end{figure}
%

We turn our attention to the case $\alpha_g=0.4$. 
The left panel of Fig.~\ref{Fig:peaklocation_alpha04} 
shows the value of $V$ as a function of  $\alpha_g\mu r_p$ 
along the line in the phase space $(\delta_r, \delta_\nu, \alpha_g\mu r_p)$
introduced in the beginning of Sec.~\ref{Sec:equilibrium_point}. 
The cases of 
$N_* =1.0$,...,$1.5$ are depicted with $0.1$ intervals.  
The right panel of Fig.~\ref{Fig:peaklocation_alpha04} 
shows the relation between $N_*$
and $\alpha_g\mu r_p$ at the equilibrium point.   
In this case,  there is only one stable point
for all values of $N_*$. Around $N_*\approx 1.2$, 
the value $\alpha_g\mu r_p$ of the equilibrium point 
rapidly decreases as $N_*$ is increased, and hence, the equilibrium point
becomes located closer to the horizon.

Our interpretation is as follows. In this axion cloud model,
the phase transition does not occur for $\alpha_g=0.4$:
The situation with two stable equilibrium points occurs 
only for $\alpha_g\lesssim 0.3$. But in our simulations
for $\alpha_g=0.4$, 
the bosenova suddenly happens. Therefore, the phase
transition seen in the $\alpha_g=0.1$ case gives a 
more correct picture. This discrepancy seems to arise because
the axion cloud model discussed here is 
a model of rough approximation.
Except for this point, however, 
the axion cloud model reproduces various characteristic 
features of the
phenomena in the simulations. For example, Fig.~\ref{Fig:peaklocation_alpha04} 
shows that as the superradiant
instability progresses, the value of $\alpha_g\mu r_p$
of the peak position becomes smaller, 
and around the ``critical'' value $N_*\approx 1.2$, 
the position rapidly becomes very close to the horizon. 
These are quite consistent with the results of our numerical simulations.

\subsection{Small oscillations around the equilibrium point}

\label{Sec:small_oscillations}

It is also interesting to study small oscillations around the
equilibrium point because it allows us to estimate the typical
dynamical time scales of the BH-axion system. Here, we introduce
the phase space parameter $q_i$ defined by
\begin{equation}
q_i = (\delta_r, \delta_\nu, \alpha_g\mu r_p).
\end{equation}
with $i=1,2,$ and $3$. The equilibrium position is denoted by
$q_i^{(0)}$ and the deviation $\Delta q_i$ from the equilibrium point
is introduced by 
$q_i = q_i^{(0)} + \Delta q_i$. Using the 
standard method in the classical mechanics, we can rewrite
the Lagrangian in terms of $\Delta q_i$ collecting 
only the second-order terms, and derive the equation
of the harmonic oscillators,
\begin{equation}
\Delta \ddot{q}_i = - \Omega_{ij} \Delta q_j.
\end{equation}
The solution of this equation can be written as a linear superposition 
of three normal modes, and their squared frequencies 
$\omega_{\rm EG}^2$
are given by the eigenvalues of the matrix $\Omega_{ij}$. 
For each value of $\omega_{\rm EG}^2$, 
there exists an eigenvector 
describing the direction of oscillation of that normal mode
in the phase space. 

In the following, we discuss the cases $N_*=1.1$ and $1.3$
for $\alpha_g=0.4$. Because the bosenova happens around
$N_*\approx 1.2$, the cases $N_*=1.1$ and $N_*=1.3$
are expected to approximate 
the state before and after the bosenova, respectively.

\subsubsection{The case $\alpha_{g}=0.4$ and $N_*=1.1$}

\label{Sec:oscillation_N1.1}

In this case, the eigenvalues of the matrix $\Omega_{ij}$
are given by 
\begin{equation}
\left(\frac{\omega_{\rm EG}}{\mu\alpha_g^2}\right)^2 
= 1.141, \ 0.249, \ 0.0166,
\end{equation}
with the corresponding eigenvectors
\begin{equation}
\Delta q_i = 
\left(\begin{array}{r}
0.110\\
-0.027\\
0.994
\end{array}
\right),
\
\left(\begin{array}{r}
0.075\\
0.724\\
0.686
\end{array}
\right),
\
\left(\begin{array}{r}
-0.378\\
-0.005\\
0.925
\end{array}
\right).
\end{equation}
Let us focus attention to the third mode with 
$\left({\omega_{\rm EG}}/{\mu\alpha_g^2}\right)^2\simeq 0.0166$.
This mode represents the oscillation
of the axion cloud in the direction of $r$
and $\delta_r$, which scarcely changes the shape in the $\nu$
direction. 
The period of the oscillation of this mode is
\begin{equation}
\Delta t \approx 761M.
\end{equation}
This is consistent with the longterm oscillation
found in the simulation (A) in Sec.~\ref{Sec:simulation_A}.
The origin of the long period of the oscillation is that the effective
potential $V(\alpha_g\mu r_p)$ becomes approximately flat 
and therefore the second-order derivative of this function
becomes very small just before
the bosenova happens.

\subsubsection{The case  $\alpha_{g}=0.4$ and $N_*=1.3$}

\label{Sec:oscillation_N1.3}

In this case, the eigenvalues of the matrix $\Omega_{ij}$
are
\begin{equation}
\left(\frac{\omega_{\rm EG}}{\mu\alpha_g^2} \right)^2
= 
14.06, \ 5.59, \ 0.175,
\end{equation}
with the eigenvectors
\begin{equation}
\Delta q = 
\left(\begin{array}{r}
0.218\\
-0.030\\
0.975
\end{array}
\right),
\
\left(\begin{array}{r}
0.070\\
0.927\\
0.367
\end{array}
\right),
\
\left(\begin{array}{r}
-0.640\\
-0.085\\
0.763
\end{array}
\right).
\end{equation}
Each eigenvalue in the case $N_*=1.1$ is larger than the corresponding
eigenvalue in the case $N_*=1.3$. The period of oscillation of 
the first mode is 
\begin{equation}
\Delta t  \approx 26M.
\end{equation}
Although this period is longer than the period $\Delta t\approx 10M$ 
observed in simulation (B) in Sec.~\ref{Sec:simulation_B}, 
this model explains that the typical dynamical time scale
after the bosenova 
becomes shorter compared to that before the bosenova.
The discrepancy would be because the axion cloud model
discussed here is a rough approximate model; 
If this model is improved so that the two stable equilibrium
positions appear also for $\alpha_g=0.4$, it would give
shorter periods of oscillations.

%
%
\section{Summary and discussion}
\label{Sec:summary}

In this paper, we have studied the 
sine-Gordon field in a Kerr spacetime motivated by 
landscape of axionlike particles/fields, i.e. the axiverse.
In order to calculate the evolution of a scalar field 
in a Kerr spacetime, we developed a 3D code that 
has the ability to describe the growth rate
by superradiant instability of the linear Klein-Gordon field
with $\lesssim 2\%$ error (Sec.~\ref{Sec:comparison}). 
Using this code, we have performed simulations 
for the scalar field mass $\alpha_g=M\mu=0.4$ and the BH rotation parameter
$a/M=0.99$ (Sec.~\ref{Sec:two_simulations}),
where the initial condition is taken to be
the quasistationary bound state of the $(\ell,m)=(1,1)$ mode 
of the Klein-Gordon equation. 
When the initial peak value is small (Sec.~\ref{Sec:simulation_A}), 
the nonlinear effect causes
periodic changes in the peak location and the peak value, and
the nonlinear effect enhances the energy and angular momentum 
extractions. On the other hand, if the initial peak value
is relatively large (Sec.~\ref{Sec:simulation_B}), the nonlinear effect
causes a collapse of the axion cloud and a subsequent
explosion, i.e. the bosenova collapse. 
During the bosenova collapse, a kind of resonance occurs
to generate waves of the $(\ell, m)=(1,-1)$ mode falling into the BH. 
Since these waves 
violate the superradiant condition, the energy flux 
toward the horizon becomes positive. Therefore, the 
energy extraction is terminated by the bosenova.

In Sec.~\ref{Sec:bosenova}, we have discussed whether
the bosenova happens or not as a result of the
superradiant instability taking account of the two possibilities, 
the bosenova collapse and 
the saturation of the superradiant instability.
We performed additional simulations
from improved initial conditions that are expected
to be more realistic.
In these simulations (the cases $C=1.08$ and $1.09$), the energy extraction
continued for a long period of time and suddenly the bosenova collapse
happened. This result supports the possibility that the
bosenova collapse actually occurs as a result of the superradiant
instability when the axion cloud
gets energy of $E\approx 1600(f_a/M_p)^2M$
(for the present setup $\alpha_g=0.4$ and $a/M=0.99$). 
If the decay
constant is order of the GUT scale, $f_a\approx 10^{16}$GeV,
the energy at the bosenova collapse is $0.16\%$ of the BH mass.

In Sec.~\ref{Sec:effective_theory}, we have discussed why the bosenova
happens by constructing an effective
theory in the nonrelativistic approximation. The axion cloud is 
assumed to be a time-dependent Gaussian wavepacket, and
the effective theory is described by dynamics of three
variables that specify its shape. The
dynamics is determined by a potential $V$ 
whose behavior depends on the amplitude of the wavepacket.  
When the value of $\alpha_g$ 
is small (e.g., $\alpha_g=0.1$ in Sec.~\ref{Sec:equilibrium_0.1}), 
there are two (outer and inner) stable minima in the potential $V$
for a small amplitude, and the outer minimum disappear
when amplitude grows to some value. Therefore,  
the bosenova collapse is explained by 
the phase transition from the outer stable point to the inner stable
point. The effective theory also indicates that
for large values of $\alpha_g$, such phase transition
does not occur and the bosenova is unlikely to happen.
The dynamical time scales observed in our simulations
before and after the bosenova also can be successfully explained
by studying small oscillations around the local minimum.
Therefore,
the axion cloud model describes the BH-axion system
fairly accurately. 
This result indicates that the critical amplitude 
for the onset of the bosenova collapse 
is primarily determined by the self-interaction 
of axions rather than the nonlinear gravity of the BH,
although the nonlinear BH gravity is important in making
the field amplitude larger by 
extraction of the rotational 
energy of the BH.

Here, we compare the bosenova phenomena in the system of BEC
atoms and in the BH-axion system. 
The action of the BH-axion system in nonrelativistic
approximation, Eq.~\eqref{Eq:action_NR}, has the same form as
the action of the BEC atoms, e.g. Eq.~(3)
of Ref.~\citen{Saito:2001-2}, except that the parabolic 
potential is replaced by the Newton potential and the 
higher-order terms are included in the nonlinear potential. 
For this reason, the two phenomena are naturally expected to 
have similarity to each other. In fact, both of the growth of the amplitude
in Fig.~\ref{Fig:amplitude_0.7} in our simulation and the implosion
of BEC atoms (e.g., Fig.~3 of Ref.~\citen{Saito:2001-2} or Fig.~1
of Ref.~\citen{Saito:2002}) are caused by nonlinear
attractive interaction. But there exists qualitative difference
between the two systems: The big implosions continue to happen 
in the BEC system, while the big implosion happens only once 
in the BH-axion system. Also, the growth of the peak height
in our system is not as sharp as the BEC system. The reason 
is that we are dealing with the $(\ell, m)=(1,1)$ mode while
the typical BEC system is in the $(\ell, m)=(0,0)$ mode. 
Since the BEC system does not have the angular momentum, 
the BEC atoms concentrate to the center  
and the peak height can become large almost unboundedly,
and this process can continue intermittently.
On the other hand, since the axion cloud is rotating
around the BH in our system, the centrifugal force
prevents the axion cloud from collapsing to the center.
For this reason, the growth of the peak height is limited.
This point is also understood by looking at the effective potential. 
In Ref.~\citen{Saito:2001-2}, an effective theory for the
BEC atoms was discussed
using a time-dependent Gaussian wave function in a similar manner to 
Sec.~\ref{Sec:effective_theory}. The effective potential
of the BEC system behaves as $f(0)=-\infty$, and this
enables the BEC atoms to concentrate at the center.
On the other hand, the effective potential 
of the BH-axion system behaves as $V(0)=\infty$,
and this makes the inner stable point. Hence,
the high concentration of axion cloud to the center is prohibited.
(Compare Fig.~1 of Ref.~\citen{Saito:2001-2} 
and Fig.~\ref{Fig:peaklocation_alpha01} of this paper.)

The authors of Ref.~\citen{Saito:2001-1,Saito:2001-2,Saito:2002} 
discussed the fact that the bursts after the implosions 
in the BEC system are caused 
by the two-body dipolar and the three-body recombination of the BEC atoms. 
The loss of atoms
causes the decrease in the attractive interaction, and the burst is
generated by the zero-point kinetic pressure. In numerical
simulations, the loss of atoms are handled by introducing the 
phenomenological terms $(-i\hbar/2)(K_2|\psi|^2+K_3|\psi|^4)\psi$
to the nonlinear Schr\"odinger equation 
(e.g., Eq.~(1) of Ref.~\citen{Saito:2002}). 
Although no such phenomenological terms are introduced in solving
the axion field in our paper, part of the axion cloud
was observed to spread out to the distant region. 
The reason would be that in the BH-axion system, 
infalling waves of the $m=-1$ mode
are generated by excitation of the bound states, causing
the loss of energy (and therefore, the attractive interaction)
of the axion cloud. 
Therefore, in our system, fall of a fraction of the axion cloud to the BH 
would play the same role as the three-body recombination of atoms
in the BEC system.

Finally, we briefly discuss whether the bosenova can be observed
by gravitational wave detectors. Studying gravitational waves
emitted from an axion cloud is rather difficult since
it requires quantum mechanical description, and the quadrupole
formula cannot be directly applied (although 
the authors of Ref.~\citen{Arvanitaki:2010}
discussed the fact that the quadrupole formula corresponds to 
level transition of axion particles). The level transition 
from the $(\ell, m)=(1,1)$ mode is prohibited by the selection rule,
and the two axion annihilation to a graviton is estimated to be
rather small \cite{Arvanitaki:2010}. 
Therefore, the gravitational wave emission
from the axion cloud scarcely affects the occurrence of the bosenova.
On the other hand, it would be possible to discuss 
gravitational waves emitted in the bosenova collapse 
by the quadrupole formula (at least in the sense of order estimate), 
since the typical time scale of the bosenova is rather long, 
$\Delta t\sim 500M$, and the nonrelativistic
approximation can be applied.

The quadrupole moment $Q_{ij}$ of the axion cloud is estimated to be
$Q_{ij}\sim r_p^2 E$, where $r_p\sim 10M$ is the position of the peak
with respect to the $r$ coordinate.
In the bosenova collapse, a burst of positive energy flux 
toward the horizon is generated, and about 5\% of the total energy 
falls into the BH in the typical time $\Delta t\sim 500M$. 
It causes decrease in $Q_{ij}$ through loss of the energy. 
Here, we ignored the shedding of the axion cloud to a far 
region because this process is slower than the burst to the horizon.
Let us consider the situation where the decay constant $f_a$
is the GUT scale, and hence, the bosenova collapse occurs
at the energy $E \approx 1.6\times 10^{-3}M$.  
Assuming the time dependence of energy as
$E=E_0+(\Delta E/2)[\cos(\pi t/\Delta t)-1]$ ($0\le t\le \Delta t$, 
where the bosenova happens at $t=0$ and ends at $t=\Delta t$)
with $\Delta E\approx 0.05E$, 
we have
\begin{equation}
\dddot{Q}_{ij}\sim r_p^2\dddot{E}
\sim r_p^2\Delta E \left(\frac{\pi}{\Delta t}\right)^3\sim
10^{-9}.
\end{equation}
From this, the energy loss rate is estimated to be
$dE/dt\sim (\dddot{Q}_{ij})^2 \sim 10^{-18}$, and the
total radiated energy is $E_{\rm rad}\sim 10^{-15}M \sim 10^{-12}E$. 
Therefore, the energy converted to gravitational waves
is expected to be very small in the bosenova. 
In a similar manner, we can estimate the
amplitude of gravitational waves emitted in this process
as 
\begin{equation}
h\sim \frac{\ddot{Q}_{ij}}{r_{\rm obs}}\sim 10^{-7}\frac{M}{r_{\rm obs}},
\end{equation}
where $r_{\rm obs}$ denotes the distance from the BH to an observer.

For the supermassive BH at the center of our galaxy, 
Sagittarius $\mathrm{A}^*$, the mass and the distance from the Sun
have been estimated to be 
$M\approx 4.5\times 10^{6}M_{\odot}$ \cite{Gillessen:2008,Ghez:2008}
and $r_{\rm obs}\approx 8\ \mathrm{kpc}$ \cite{Ghez:2008,Eisenhauer:2003}.
For these values, we have
$M/r_{\rm obs}\sim 10^{-11}$, and therefore, $h\sim 10^{-18}$. 
The frequency of gravitational waves emitted in this process is
$\sim 1/\Delta t\sim 10^{-4}\ \mathrm{Hz}$. 
The strain amplitude $h_{\rm rss}:= [\int |h|^2dt]^{1/2}$ 
of the gravitational wave burst in this process is 
$h_{\rm rss}\sim 10^{-16}(\mathrm{Hz})^{-1/2}$, 
and for the frequency $10^{-4}\ \mathrm{Hz}$, 
this value is (by order one) above the 
threshold of the sensitivity of the future-planned 
space-based gravitational wave detector, the LISA \cite{GW_detectors}. 
For the BH candidate Cygnus X-1, 
the mass and the distance have been estimated to be 
$M\approx 8.7\pm 0.8 M_{\odot}$ \cite{Shaposhnikov:2007} 
and 
$r_{\rm obs}\approx 1.86^{+0.12}_{-0.11}\ \mathrm{kpc}$ \cite{Reid:2011}.
For these values, we have $M/r_{\rm obs}\sim 10^{-16}$,
and therefore, $h\sim 10^{-23}$. The frequency of gravitational waves
emitted in this process is $\sim 100\ \mathrm{Hz}$.
The strain amplitude is $h_{\rm rss}\sim 10^{-24}(\mathrm{Hz})^{-1/2}$,
and for the frequency $100\ \mathrm{Hz}$, this value is below the 
threshold of the sensitivity of the planned ground-based
gravitational wave detectors, the Advanced LIGO, 
the Advanced Virgo,
and the LCGT \cite{GW_detectors}.

Note that the estimate here has been done for the 
parameters $\alpha_g=M\mu=0.4$, $a/M=0.99$, and 
$f_a=10^{16}$GeV. For other parameters, e.g. if the
value of $f_a$ is smaller, the detection of
gravitational waves from the bosenova collapse is more difficult.
However, if the value of $f_a$ is around the GUT scale or
somewhat larger, 
we have the possibility of detection of gravitational wave bursts 
from the bosenova. Although we have discussed only bursts here,
it is also interesting to study gravitational waves that originate
from the oscillation of the axion cloud during the bosenova
whose period is $\sim 10M$. Since this oscillation continues
at least for $\sim 1000M$, the detection may be more plausible. 
Also, it is interesting to take account of 
the possibility of existence of unknown BHs in the neighborhood
of the Sun, because all existing candidates for stellar mass BHs
are X-ray binaries while many isolated BHs that cannot be seen by
electromagnetic waves are expected to exist.

The detailed studies on the gravitational wave emission
in the BH-axion system during the bosenova are necessary 
in order to improve the above rough 
estimate and to obtain indication 
of the existence of axionlike particles or 
constraints on them.
Another interesting observational possibility is that 
if the BH is immersed in the magnetic field
and the axion field $\Phi$ has coupling to the electromagnetic field
through the Chern-Simons interaction $\mathcal{L}_{a\gamma\gamma}= g_{a\gamma\gamma}
\Phi\boldsymbol{E}\cdot\boldsymbol{B}$, the axion cloud may 
radiate electromagnetic waves. Studying the feature 
of the electromagnetic radiation in this process and
exploring the possibility of observing this phenomena are also
interesting issues to be investigated.

\section*{Acknowledgements}
H.Y. thanks Hajime Sotani, Hisa-aki Shinkai 
and Kunihito Ioka for helpful comments. 
This work was supported by the Grant-in-Aid for
Scientific Research (A) (22244030).

\appendix

%
%
\section{Green's function analysis}
\label{Sec:Green}

%
\begin{figure}[tb]
\centering
{
\includegraphics[width=0.45\textwidth]{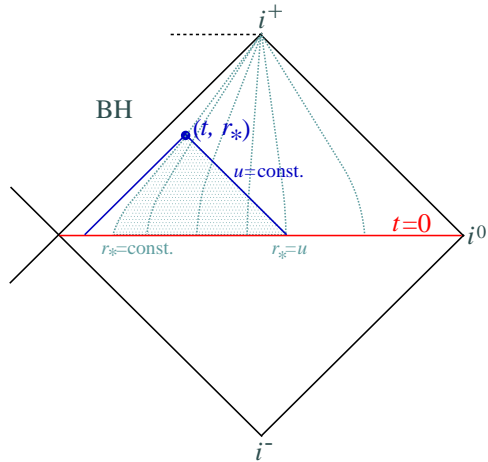}
}
\caption{The domain of integration shown in the Penrose
diagram of a Kerr spacetime.}
\label{Fig:Kerr_Penrose}
\end{figure}
%

In this appendix, we study which modes are generated
by the nonlinear effect when the amplitude of 
$\varphi$ is relatively small using a perturbative approach
with the Green's function method. 
In particular, we pay attention to whether 
waves of $(\ell,m)=(1,-1)$ mode found in
our simulations can be generated. 
We decompose the field as 
\begin{equation}
\varphi(x) = \varphi_0(x) + \Delta\varphi,
\end{equation}
where $\varphi_0$ is the bound state of the $(\ell,m)=(1,1)$ mode
of the Klein-Gordon equation:
\begin{equation}
\varphi_0 =2\mathrm{Re}\left[ 
e^{(\gamma - i\omega_0)t} P(r)S_1^1(\cos\theta)e^{i\phi}
\right],
\end{equation}
and $\Delta\varphi$ denotes the deviation generated 
by the nonlinear effect. In the following, we consider
the situation where $\varphi_0$ is relatively small,
and take account of the order up to $O(\varphi_0^3)$ and
ignore terms of $O(\varphi_0^4)$. Then, the sine-Gordon equation
is approximated as
\begin{equation}
(\Box-\mu^2)\Delta\varphi =J,
\label{Eq:perturbative_equation}
\end{equation}
with 
\begin{equation}
J:=-\frac{\mu^2}{6}\varphi_0^3
\end{equation}
Equation~\eqref{Eq:perturbative_equation} is a linear
equation with a source term $J$. In order to solve this type
of equation, the Green's function method is useful. The Green's
function is defined by 
\begin{equation}
(\Box^\prime-\mu^2) G(x,x^\prime)
=\delta^4(x,x^\prime),
\end{equation}
where $x$ and $x^\prime$ denote some points in the spacetime,
and hereafter, prime ($\prime$) indicates the coordinates $x^\prime$.
Assuming the initial condition at $t=0$ to be
$\Delta\varphi = \partial_t(\Delta\varphi)=0$, the solution
of $\Delta\varphi$ is written in terms of the Green's function as
\begin{equation}
\Delta\varphi = 
\int_{D^\prime}d^4x^\prime\sqrt{-g(x^\prime)}G(x,x^\prime)J(x^\prime).
\label{Eq:formal_solution}
\end{equation}
Here, for $x=(t,r_*,\theta,\phi)$ and 
$x^\prime=(t^\prime,r_*^\prime,\theta^\prime,\phi^\prime)$, 
the domain $D^\prime$ is taken as  
the triangular region $u^\prime \le u$, $v^\prime \le v$, and $t^\prime\ge 0$,
where $u$ and $v$ are null coordinates, 
$u=t+r_*$ and $v=t-r_*$ (see Fig.~\ref{Fig:Kerr_Penrose}).

The Green's function can be constructed in terms of 
the eigenfunctions of the operator $\Box-\mu^2$. 
The specific form is  
\begin{equation}
G(x,x^\prime) = 
\frac{1}{(2\pi)^2}\sum_{\ell, m}\int_{-\infty}^{\infty}d\omega
G_{\ell m}^{\omega}(r,r^\prime)
e^{-i\omega(t-t^\prime)+im(\phi-\phi^\prime)}
S_\ell^m(\cos\theta)\bar{S}_\ell^m(\cos\theta^\prime),
\label{Eq:Green_function}
\end{equation}
where bar denotes the complex conjugate and 
\begin{equation}
G_{\ell m}^{\omega}(r,r^\prime) = 
\frac{1}{W_{\ell m\omega}}\left[
\theta(r-r^\prime)R^+_{\ell m\omega}(r)R^-_{\ell m\omega}(r^\prime)
+
\theta(r^\prime-r)R^-_{\ell m\omega}(r)R^+_{\ell m\omega}(r^\prime)
\right].
\end{equation}
Here, $R_+$ and $R_-$ are radial functions
satisfying the boundary conditions
\begin{equation}
R^+_{\ell m \omega} \simeq \begin{cases}
\quad e^{ikr}/r, & r\to \infty;\\
A^{[+]\omega}_{\ell m}e^{i\tilde{\omega}r_*}+
B^{[+]\omega}_{\ell m}e^{-i\tilde{\omega}r_*}, & r\simeq r_+,
\end{cases}
\label{Eq:R+mode}
\end{equation}
\begin{equation}
R^-_{\ell m \omega} \simeq \begin{cases}
A^{[-]\omega}_{\ell m}e^{-ikr}/r+
B^{[-]\omega}_{\ell m}e^{ikr}/r, & r\to \infty;\\
\quad e^{-i\tilde{\omega}r_*}, & r\simeq r_+,\\
\end{cases}
\end{equation}
where $k = \sqrt{\omega^2-\mu^2}$ and we impose $\mathrm{Im}[k] \ge 0$.
The function $R^+$ is the solution satisfying the outgoing/decaying 
boundary condition at infinity, and the function $R^-$
is the solution satisfying the ingoing boundary condition at the horizon.
Choosing $R^+$ and $R^-$ in this way, the solution $\Delta\varphi$
satisfies both of the ingoing boundary condition at the horizon
and the outgoing/decaying condition at infinity. 
$W_{\ell m\omega}$ is the Wronskian determined by
\begin{equation}
W_{\ell m\omega}(R^-,R^+):=
\Delta\left(-R^+\partial_rR^{-}+R^-\partial_rR^{+}\right).
\end{equation}
The value of $W_{\ell m\omega}$ is constant for arbitrary $r$,
and it is calculated as
\begin{equation}
W_{\ell m\omega}(R^-, R^+) = 2i\tilde{\omega} (r_+^2+a^2) A^{[+]\omega}_{\ell m}
= 2ik A^{[-]\omega}_{\ell m}. 
\end{equation}
Since we do not know the analytic expressions of $R^-$ and $R^+$ in the 
whole region $r_+\le r<\infty$, 
we cannot derive the exact solution. However, we can extract various
properties of the solution of $\Delta \varphi$.

Substituting the Green's function \eqref{Eq:Green_function}
into Eq.~\eqref{Eq:formal_solution}, we have the following formula:
\begin{equation}
\Delta\varphi = 
\sum_{\ell, m}
\int_{-\infty}^{\infty}d\omega
e^{-i\omega t + im\phi}
S_{\ell}^{m}(\cos\theta)
\left[R^+_{\ell m \omega}(r)X^+_{\ell m\omega}(t,r)
+R^-_{\ell m\omega}(r)X^{-}_{\ell m\omega}(t,r)
\right],
\label{Eq:Deltaphi_formal}
\end{equation}
where $X^+_{\ell m\omega}$ and $X^-_{\ell m\omega}$ are defined by
\begin{equation}
X^+_{\ell m\omega}(t,r) = 
\frac{1}{W_{\ell m \omega}}
\int_{-v}^{r_*}dr_*^\prime
\frac{\Delta^\prime}{r^{\prime 2}+a^2}
\int_{0}^{v+r_*^\prime}
dt^\prime e^{i\omega t^\prime} R^-_{\ell m\omega}(r^\prime)
J_{\ell m\omega}(t^\prime, r^\prime),
\end{equation}
\begin{equation}
X^-_{\ell m\omega}(t,r) = 
\frac{1}{W_{\ell m \omega}}
\int_{r_*}^{u}dr_*^\prime
\frac{\Delta^\prime}{r^{\prime 2}+a^2}
\int_{0}^{u-r_*^\prime}
dt^\prime e^{i\omega t^\prime} R^+_{\ell m\omega}(r^\prime)
J_{\ell m\omega}(t^\prime, r^\prime),
\end{equation}
with 
\begin{equation}
J_{\ell m\omega} =
-\frac{1}{(2\pi)^2}\int_0^{2\pi}d\phi
\int_{-1}^{1}d\nu (r^2+a^2\nu^2)e^{-im\phi}S_\ell^m(\nu)
U^\prime_{\rm NL}(\varphi_0),
\end{equation}
where $\nu:=\cos\theta$. 
Here, we introduce further approximation: we replace the spheroidal
harmonics $S_{\ell}^me^{im\phi}$ by
the spherical harmonics $P_{\ell}^{m}e^{im\phi}$.
Although this approximation holds only when $|a^2k^2|$ is small,
this formula is assumed for all values of $\omega$.
In this approximation, $J_{\ell m\omega}$ becomes independent
of $\omega$, and therefore, we simply denote it by $J_{\ell m}$. 
Then, the integration can be performed as
\begin{equation}
J_{\ell m} =
K_{\ell}^m(r)e^{(3\gamma-im\omega_0 t)}P(r)^{\frac{3+m}{2}}\bar{P}(r)^{\frac{3-m}{2}}.
\end{equation}
where
\begin{subequations}
\begin{eqnarray}
K_{3}^{\pm 3}&=&\frac{\mu^2}{12\sqrt{210}\pi}(9r^2+a^2),
\\
K_{3}^{\pm 1}&=& \frac{\mu^2}{20\sqrt{14}\pi}(3r^2-a^2),
\\
K_{1}^{\pm 1}&=& \frac{3\mu^2}{140\pi}(7r^2+a^2),
\end{eqnarray}
\end{subequations}
and $K_\ell^m=0$ for the other values of $\ell$ and $m$.

Now, let us consider the point $(t,r)$ in the neighborhood
of the horizon, i.e., $r\simeq r_+$.
For this point, it can be directly checked that 
$X^+_{\ell m\omega}(t,r)\simeq 0$, 
and therefore, $\Delta\varphi$ satisfies the ingoing boundary
condition at the horizon. On the other hand, $X^{-}_{\ell m\omega}(t,r)$
becomes nonzero and can be written 
\begin{multline}
X^-_{\ell m\omega}(t,r) = 
\frac{1}{W_{\ell m \omega}}
\int_{r_*}^{u}dr_*^\prime
\frac{\Delta^\prime}{r^{\prime 2}+a^2}
\frac{e^{[3\gamma+i(\omega-m\omega_0)](u-r_*^\prime)}-1}{3\gamma+i(\omega-m\omega_0)}
\\
\times
R^+_{\ell m\omega}(r^\prime)
K_{\ell}^m(r^\prime)P(r^\prime)^{\frac{3+m}{2}}\bar{P}(r^\prime)^{\frac{3-m}{2}}.
\end{multline} 
Substituting this formula into Eq.~\eqref{Eq:Deltaphi_formal},
we get
\begin{multline}
\Delta\varphi 
=
\sum_{\ell, m}
e^{im\phi}
S_{\ell}^{m}(\cos\theta)
\frac{e^{im\Omega_Hr_*}}{2i(r_+^2+a^2)}
\\
\times
\left\{
e^{(3\gamma-im\omega_0)u}D_{\ell m}(u,r_*)
-
\int_{-\infty}^{\infty}d\omega
\frac{e^{-i\omega u}}{\tilde{\omega}A^{[+]\omega}_{\ell m}
[3\gamma+i(\omega-m\omega_0)]}
E_{\ell m}^{(\omega)}(u,r_*)
\right\},
\label{Eq:Deltaphi_solution_last}
\end{multline}
with 
\begin{multline}
D_{\ell m}(u,r_*) = 
\int_{-\infty}^{\infty} 
d\omega
\frac{1}{\tilde{\omega}A^{[+]\omega}_{\ell m}
[3\gamma+i(\omega-m\omega_0)]}
\\
\times
\int_{r_*}^{u}dr^\prime_*
\frac{\Delta^\prime}{r^{\prime 2}+a^2}
e^{-[3\gamma+i(\omega-m\omega_0)]r_*^\prime}
R^+_{\ell m\omega}(r^\prime)K_\ell^m(r^\prime)
P(r^\prime)^{\frac{3+m}{2}}\bar{P}(r^\prime)^{\frac{3-m}{2}}
\end{multline}
and
\begin{equation}
E^{(\omega)}_{\ell m}(u,r_*) = 
\int_{r_*}^{u}dr^\prime_*
\frac{\Delta^\prime}{r^{\prime 2}+a^2}
R^+_{\ell m\omega}(r^\prime)K_\ell^m(r^\prime)
P(r^\prime)^{\frac{3+m}{2}}\bar{P}(r^\prime)^{\frac{3-m}{2}}.
\end{equation}
Here, $D_{\ell m}(u,r_*)$ and $E_{\ell m}(u,r_*)$ are
slowly varying functions with respect to $u$ and $r_*$ for 
$r_*\ll -M$ and $u\gg M$.

Let us consider the first term of Eq.~\eqref{Eq:Deltaphi_solution_last}.
The dependence of the first term on $t$ and $\phi$ is 
$\sim e^{3\gamma t+im(\phi-\omega_0t)}$, and this is a mode propagating
to $+\phi$ direction.  
The integration of the second term is performed using the
techniques of the complex analysis.
Defining the contour $C$ in a complex $\omega$ plane
that goes along the real line from $-\infty$ to $\infty$
and then clockwise along a semicircle centered at zero in the
lower-half plane, the integral is rewritten as the sum of
contribution from the poles and 
the integration along the branch cut. Since the branch cut 
integral typically gives a subdominant contribution, 
we focus attention to the poles.
The singular points of the integrand appear at $\omega=m\Omega_H$,
$\pm\mu$, $m\omega_0+3\gamma$, and $\omega_{\rm BS}^{(\ell m n)}$ 
satisfying $A_{\ell m}^{[+]\omega_{\rm BS}^{(\ell m n)}}=0$ $(n=1,2,3,...)$. 
Among them, $\omega=m\Omega_H$ and $\pm\mu$ become endpoints
of the branch cut, and therefore, they are not poles. 
By applying the residue theorem to the poles 
$\omega=m\omega_0+3i\gamma$ and $\omega_{\rm BS}^{(\ell m n)}$, 
the terms proportional to $e^{3\gamma t + im(\phi-\omega_0 t)}$ and 
$e^{i(m\phi-\omega_{\rm BS}^{(\ell m n)}t)}$ appear, respectively.
Among these two, $e^{3\gamma t + im(\phi-\omega_0 t)}$ represents
waves propagating in the $+\phi$ direction.

In order to understand the behavior of 
$e^{i(m\phi-\omega_{\rm BS}^{(\ell m n)}t)}$, we have to 
evaluate $\omega_{\rm BS}^{(\ell m n)}$. 
From Eq.~\eqref{Eq:R+mode}, 
the condition $A_{\ell m}^{[+]\omega_{\rm BS}^{(\ell m n)}}=0$ gives
the mode satisfying the ingoing boundary condition at the
horizon and the decaying/outgoing boundary condition at infinity
simultaneously. This is the bound state discussed in 
Sec.~\ref{Sec:Superradiant_instability}. The typical 
mode of the bound state satisfies
$\omega_{\rm BS}^{(\ell m n)2}\simeq \mu^2\simeq\omega_0^2$,
since the gravitational binding energy is not so large. Then,
typical value of $\omega_{\rm BS}$ is estimated as
\begin{equation}
\omega_{\rm BS}^{(\ell m n)}\simeq \pm \omega_0.
\end{equation}
Remember that the poles for the bound states typically appear
in both right and left half complex planes. 
If we adopt $m=1$, the behavior of $e^{i(m\phi-\omega_{\rm BS}^{(\ell m n)}t)}$
becomes $e^{i(\phi\mp\omega_{0}t)}$, and the waves of 
negative frequency $\omega_{\rm BS}^{(\ell m n)}\simeq -\omega_0$
propagate to the $-\phi$ direction 
(i.e., they are waves of the $m=-1$ mode).



\end{document}